\documentclass[twocolumn,prx,longbibliography]{revtex4-2}
\usepackage{graphicx}
\usepackage{amsmath}
\usepackage{amsfonts}
\usepackage{mathtools}
\usepackage{subfigure}
\usepackage{mhchem}

\begin{document}
\newcommand{\nwc}{\newcommand}
\nwc{\vs}{\vspace}
\nwc{\hs}{\hspace}
\nwc{\la}{\langle}
\nwc{\ra}{\rangle}
\nwc{\lw}{\linewidth}
\nwc{\nn}{\nonumber}
\nwc{\pd}[2]{\frac{\partial #1}{\partial #2}}
\newcommand{\be}{\begin{equation}}
\newcommand{\ee}{\end{equation}}
\newcommand{\bea}{\begin{eqnarray}}
\newcommand{\eea}{\end{eqnarray}}
\def \hh{\mathcal{I}}
\def \pr{\textrm{Prob}}
\def\A{\mathcal{A}}
\renewcommand{\vec}{\boldsymbol}

\title{Second law for active heat engines}

\author{Arya Datta$^1$,  Patrick Pietzonka$^2$ and,  Andre C Barato$^1$}
\affiliation{$^1$Department of Physics, University of Houston, Houston, Texas 77204, USA\\
$^2$Max Planck Institute for the Physics of Complex Systems, N\"othnitzer Stra{\ss}e 38, 01187 Dresden, Germany}

\parskip 1mm
\def\d{{\rm d}}
\def\Ps{{P_{\scriptscriptstyle \hspace{-0.3mm} s}}}
\def\MF{{\mbox{\tiny \rm \hspace{-0.3mm} MF}}}
\begin{abstract}
Macroscopic cyclic heat engines have been a major motivation for the emergence of thermodynamics. In the last decade, cyclic heat engines that have large fluctuations and operate 
at finite time were studied within the more modern framework of stochastic thermodynamics. The second law for such heat engines states that the efficiency cannot be larger than the Carnot efficiency.
The concept of cyclic active heat engines for a system in the presence of hidden 
dissipative degrees of freedom, also known as a nonequilibrium or active reservoir, has also been studied in theory and experiment. Such active engines show rather interesting 
behavior such as an ``efficiency'' larger than the Carnot bound. They are also likely to play an important role in future developments,  given the ubiquitous presence of 
active media. However, a general second law for cyclic active heat engines has been lacking so far. Here, by using a known inequality in stochastic
thermodynamics for the excess entropy, we obtain a general second law for active heat engines, which does not involve the energy dissipation
of the hidden degrees of freedom and is expressed in terms of quantities that can be measured directly from the observable degrees
of freedom. Besides heat and work, our second law contains an information-theoretic term, which allows an active heat engine to
extract work beyond the limits valid for a passive heat engine. To obtain a second law expressed in terms of observable variables in the presence 
of hidden degrees of freedom we introduce a coarse-grained excess entropy and prove a fluctuation theorem for this quantity. 
\end{abstract}

\maketitle

\section{Introduction}

Thermodynamics \cite{callen} is a major theory in physics that deals with the transfer  of heat into other forms of energy. 
This theory provides universal bounds through its second law on processes that involve heat transfer.
Cyclic engines that transform part of the heat taken from a hot reservoir into useful work are central objects in thermodynamics. 
One of the most prominent statements of the second law of thermodynamics is the Carnot bound on the efficiency of heat engines. 
Yet standard thermodynamics has considerable limitations: only heat engines with negligible fluctuations and that operate in the quasi-static limit 
can be analyzed quantitatively. 

Real engines, however, operate at finite time. Furthermore, it is now possible to realize in experiments heat engines made of a small number of constituents, 
for which fluctuations cannot be neglected. The modern theory of stochastic thermodynamics \cite{oono98,seif12} does not suffer from the same limitations as standard thermodynamics and is able 
to deal with such heat engines. Within stochastic thermodynamics a model for a cyclic heat engine with a colloidal particle in
a harmonic potential has been proposed in \cite{schm08}. This model has been realized in an experiment in \cite{blic12}. Much theoretical \cite{espo10,izum11,tu14,bran15,raz16,ray17,koyu19}
and experimental \cite{stee11,mart15,mart16,ross16,mart17} work  has been done on such engines. For such heat engines, the Carnot bound on efficiency remains valid.

A novel perspective was put forward in an experiment of a so-called active heat engine \cite{kris16}. This cyclic heat engine is likewise composed of a 
colloidal particle in a harmonic trap. However, in contrast to a passive heat engine the colloidal particle is not immersed in  a 
solution that simply acts as an equilibrium heat bath. Rather the solution has bacteria that collide with the particle. Such a bacterial bath may be called 
an nonequilibrium or active reservoir. Key observations made in this work were that an active heat engine can extract more work than its passive counter-part and that the 
pseudo-efficiency (work divided by heat taken from hot reservoir) of the engine can be larger than the Carnot efficiency, which might be called a ``violation'' of  the second law. 

Theoretical work on cyclic active heat engines has been prompted by this experiment \cite{zaki17,saha19,ekeh20,holu20,holu20b,kuma20,lee20,fodo21,gron21}. Furthermore, a general theoretical 
framework for steady state or autonomous active engines has been proposed in \cite{piet19}. Overall, active matter \cite{rama10,marc13,bech16,juli18} is a novel key concept in nonequilibrium statistical 
mechanics and there has been considerable interest in the stochastic thermodynamics of systems immersed in active external media \cite{argu16,piet17c,mand17,spec18,shan18,dabe19,gopa21,mark21}.

A standard explanation that can be given to the observation of a pseudo-efficiency larger than the Carnot efficiency is that if we properly account for the energy dissipated 
by the bacteria then there is no violation of the second law. This explanation is in principle correct, if we also consider the dissipation of the bacteria then 
the total entropy production must be positive. However, such statement of the second law is rather limited. First, the energy dissipation of the bacteria is probably 
much larger than the extracted work and would be the dominating term is such second law. Hence, this second law would not represent a meaningful bound on the amount 
of extracted work. Second, this energy dissipation is related to degrees of freedom that are not accessible in the experiment since only the position of the colloidal 
particle particle is measured.

Beyond this particular experiment, an active heat engine is an interesting concept. We can imagine a cyclic heat engine with a system composed of a few degrees of freedom. 
This system interacts with, possibly hidden, dissipative degrees  freedom that make the external medium active. A second law that contains the 
energy dissipation due to this activity  has two main issues. First, the energy dissipated per cycle period  due to the activity  increases indefinitely if we increase the cycle period, 
since a nonequilibrium system dissipates energy even after reaching a steady state. However, the extracted work per period saturates and becomes a constant 
for large period. Second, this energy dissipation might be related to degrees of freedom that cannot be observed in an experiment and their measurement is not 
necessary to calculate  extracted work. 

What is the correct statement of the second law for active heat engines? Such statement must fulfill the following requirements. 
First, the bound on extracted work can only contain terms that have the same scaling with the period of the engine. Second, it must accommodate the observations made in the experiment 
from \cite{kris16}, such as a pseudo-efficiency larger than the Carnot efficiency. Third, for a passive heat engine,  it must become the known standard second law. 
Fourth, it must contain terms that depend only on the observable degrees of freedom.

In this paper we show that the appropriate statement of the second law for generic cyclic active heat engines, which fulfills all the aforementioned requirements, 
is obtained  from a well known quantity in stochastic thermodynamics called excess entropy \cite{hata01,trep04,spec05a,cher06,pere11,saga11b,seif12} (or the related 
non-adiabatic entropy \cite{espo10b,espo10d,espo10f}). Crucially, we obtain this result with a suitable decomposition of the excess entropy. Beyond the known 
excess entropy, in order to fulfill the fourth requirement, we introduce a coarse-grained excess entropy and prove a fluctuation theorem for it.

This decomposition of the excess entropy for cyclic active heat engines leads to three terms, the heat taken from the hot reservoir, 
the extracted work and a novel term that has an  information theoretical interpretation. Interestingly, this term relates the non-equilibrium steady state distribution associated with 
the active heat engine to the equilibrium Boltzmann distribution associated with its passive counterpart. Our results are illustrated with three  
models that allow for  analytical calculations and one numerical model inspired by the experiment in \cite{kris16}. 

The paper is organized as follows. In Sec. \ref{sec2} we use a two-state model to illustrate our main result. 
The general formalism of our second law for active heat engines is introduced in Sec. \ref{sec3}. In Sec. \ref{sec4} we introduce the coarse-grained second law, 
which is expressed in terms of quantities that only depend on an observable degree of freedom. In Sec. \ref{sec5} we discuss three simple models for active heat engines that allow for 
analytical calculations. Sec. \ref{secnew} contains a more complex model that is analyzed with numerics. We conclude in Sec. \ref{sec6}. 
The fluctuation theorem for the coarse-grained excess entropy is proved in the appendix.

\section{Second law for two-state heat engine}
\label{sec2}

\begin{figure}

\subfigure[]{\includegraphics[width=10cm]{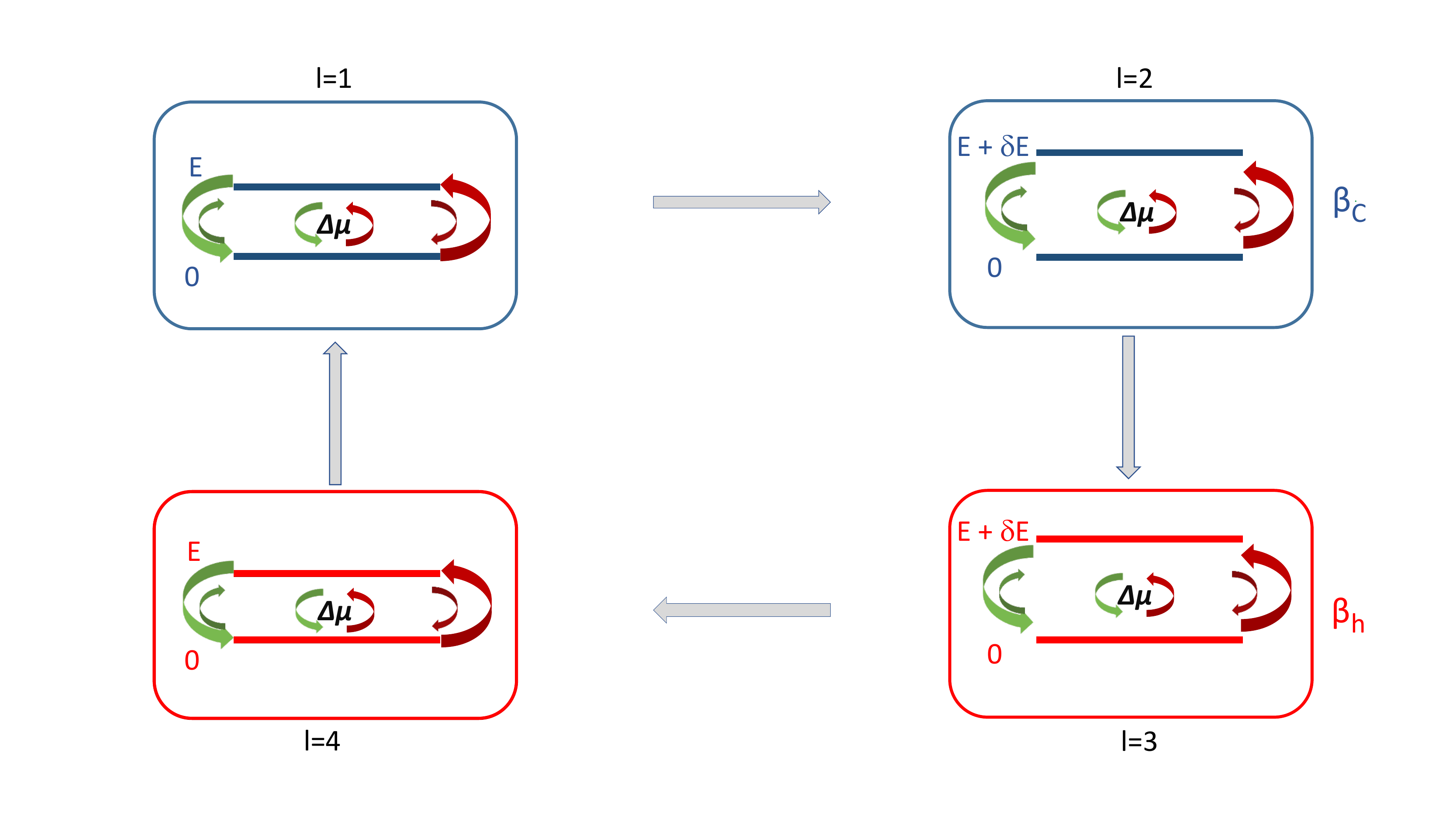}\label{fig1a}}
\subfigure[]{\includegraphics[width=8cm]{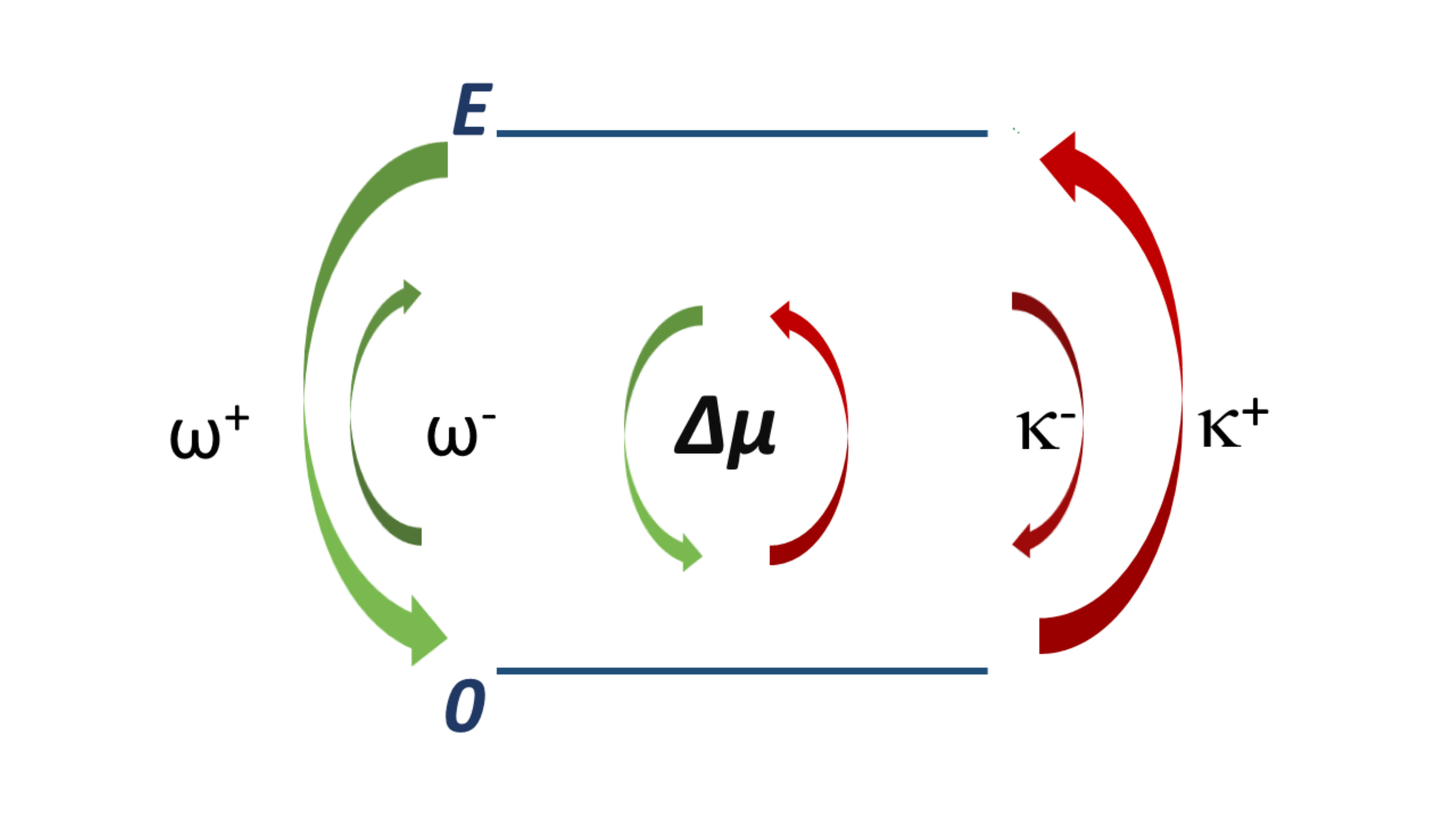}\label{fig1b}}
\vspace{-2mm}
\caption{(a) Depiction of the protocol for the two-state heat engine with a period divided into four parts labeled by $l$. The four changes are as follows: from $l=1$ to $l=2$ the 
energy changes from   $E$ to $E+\delta E$, from $l=2$ to $l=3$ the inverse temperature changes from $\beta_c$ to $\beta_h$, from $l=3$ to $l=4$ the energy changes from 
from $E+\delta E$ to $E$, and from $l=4$ to $l=1$ the inverse temperature changes from $\beta_h$ to $\beta_c$.  (b) Possible reactions for the two-state system.
The transition rate for the forward (backward) phosphorylation reaction in Eq. \eqref{eqphos} is $\kappa_+$ ($\kappa_-$) and the transition rate for the forward (backward) 
dephosphorylation reaction in Eq. \eqref{eqdephos} is $\omega_+$ ($\omega_-$). We parametrize the transition rates as $\kappa^+(t)=\omega^-(t)=k$, $\kappa^-(t)=k\textrm{e}^{\beta(t)[E(t)+\Delta\mu]}$, and 
$\omega^+(t)= k\textrm{e}^{\beta(t)E(t)}$.
}   
\label{fig1} 
\end{figure}

Consider the two-state heat engine shown in Fig. \ref{fig1}. It has a down state with energy zero and an up state with energy larger than zero. The protocol of this engine with period $\tau$  is as follows. 
It has four parts, each with duration $\tau_l$, with $\tau_1+\tau_2+\tau_3+\tau_4=\tau$. There are two energy changes: at time $\tau_1$ the energy changes from $E$ to $E+\delta E$ and at 
time $\tau_1+\tau_2+\tau_3$ the energy changes back from  $E+\delta E$ to $E$. Moreover, there are two temperature changes: at time $\tau_1+\tau_2$ the temperature changes from $\beta_c$ to $\beta_h$ and at 
time $\tau$ the temperature changes back from  $\beta_h$ to $\beta_c$.


Work can be extracted in this cycle in the following way. If the system finishes the third part at the up state then an amount $\delta E$ of work is extracted. Moreover, 
if the system finishes the first part at the up state then an amount $\delta E$ of work is spent on the system ($-\delta E$ of extracted work). Hence, the maximum  
amount of work that can be extracted in a cycle is $\delta E$.

A similar passive two-state heat engine has been studied in \cite{espo10}. The known second law for this passive heat engine reads 
\begin{equation}
(\beta_c-\beta_h)Q_h-\beta_cW\ge 0,
\label{eqseclaw1}
\end{equation}
where $Q_h$ is the average heat taken from the hot reservoir during a period and $W$ is the average extracted work during a period. This second law implies the well 
known bound on the efficiency of a passive engine, 
\begin{equation}
W/Q_h \le\eta_C\equiv 1-\beta_h/\beta_c,
\end{equation}
where $\eta_C$ is the Carnot efficiency.

Now we consider an active two-state heat engine. The simplest way to make the engine active is to include two different transition paths between the two states. 
A possible physical interpretation shown in Fig. \ref{fig1b} 
would be an enzyme that can be in two states, $X$ and $XP$, where $XP$ is the phosphorylated form of the enzyme. The state $X$ has energy $0$ and the state $XP$ has non-zero energy, which can be either $E$
or $E+\delta E$. One transition path  from $X$ to $XP$ is a phosphorylation reaction 
\begin{equation}
X+ATP\rightleftharpoons XP+ADP.
\label{eqphos}
\end{equation}
The other transition path is a dephosphorylation reaction 
\begin{equation}
XP\rightleftharpoons X+P.
\label{eqdephos}
\end{equation}
During any of the four steps of the cycle the enzyme burns ATP. The free energy of one ATP hydrolysis is written  as $\Delta\mu=\mu_{ATP}-\mu_{ADP}-\mu_{P}$, where $\mu$ is the chemical potential. 
This exactly solvable model for an active heat engine is fully defined in Sec. \ref{sec52}. In this section we use the results obtained with this model for illustrative purposes. 

\begin{figure}
\subfigure[]{\includegraphics[width=8cm]{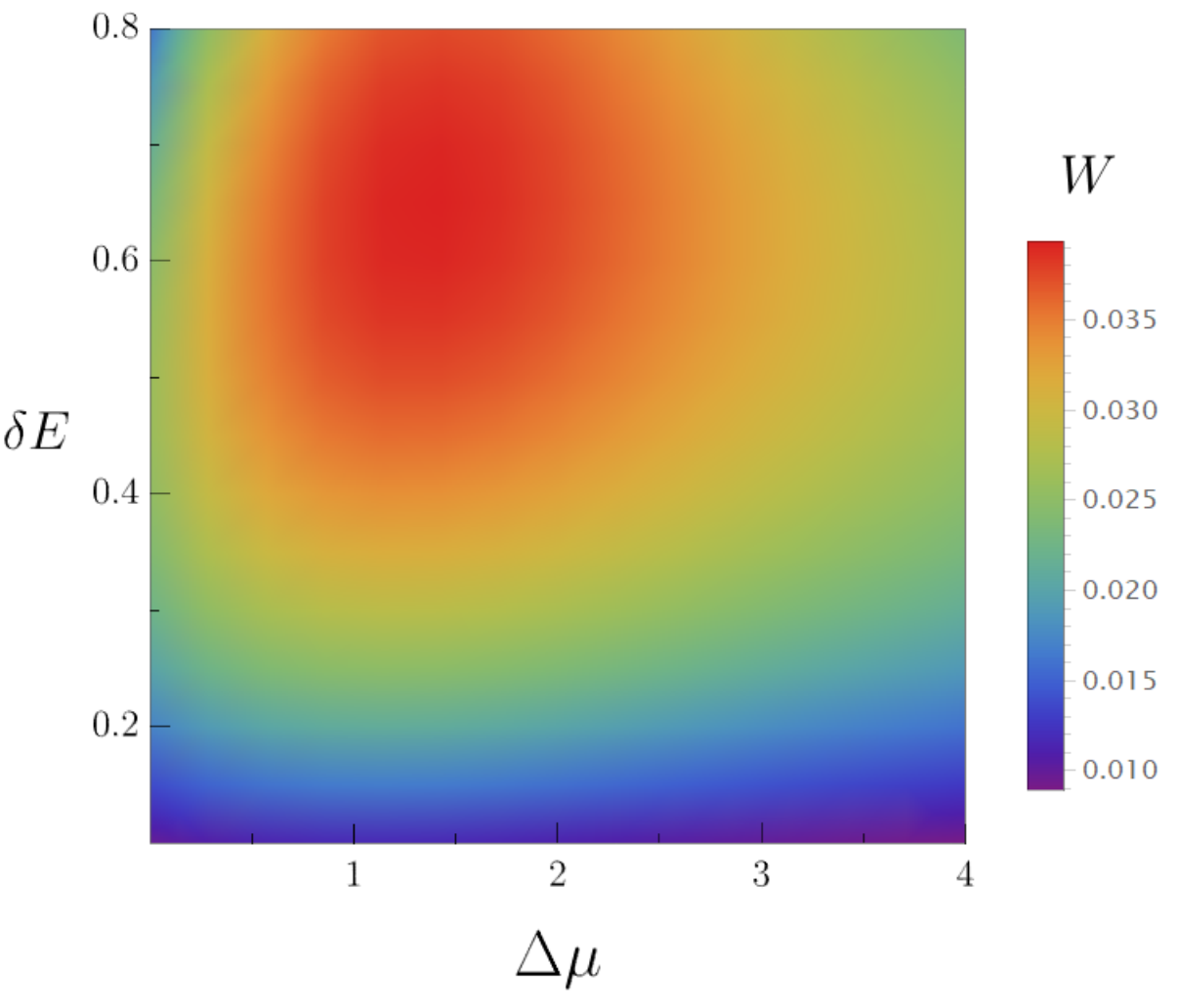}\label{fig2a}}
\subfigure[]{\includegraphics[width=8cm]{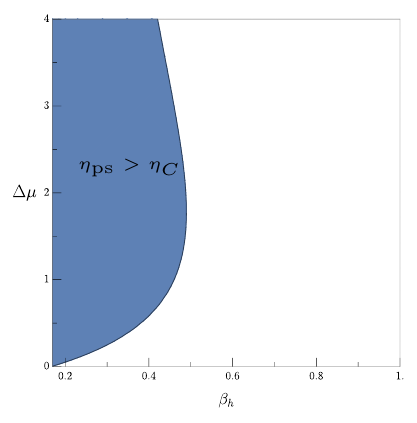}\label{fig2b}}
\vspace{-2mm}
\caption{Work and efficiency for the two-state heat engine depicted in Fig. \ref{fig1}.
(a) Work as a function of $\Delta\mu$ and $\delta E$  for $E=1$ and $\beta_h=0.5$. Work for an active heat engine can be larger as compared to the 
work of its passive counterpart for $\Delta \mu=0$. (b) The pseudo-efficiency $\eta_{\textrm{ps}}$ is larger than the Carnot efficiency $\eta_C$ in the blue filled 
region, for $E=0.1$ and $\delta E=0.5$. For both figures $\beta_c=1$, $k=1$, and $\tau=1$. These parameters correspond to an arbitrary cold temperature. 
The hot temprature is the cold temperature times $\beta_h^{-1}$. The parmaeters $E$, $\delta E$, and $\Delta \mu$ are all in units of $\beta_c^{-1}$.
}   
\label{fig2} 
\end{figure}

If $\Delta\mu=0$ the heat engine is a standard passive heat engine and the second law in Eq. \eqref{eqseclaw1} applies. For $\Delta\mu\neq 0$ the heat engine is an active heat engine. 
As shown in Fig. \ref{fig2}, this simple active heat engine produces remarkable behavior. First, making the engine active with a non-zero $\Delta\mu$ can increase the 
amount of extracted work. Second, the pseudo-efficiency 
\begin{equation}
\eta_{\textrm{ps}}\equiv W/Q_h 
\end{equation}
can be larger than the Carnot efficiency $\eta_C$. The heat $Q_h$ requires a word of caution. This quantity is not exactly the heat taken from the hot reservoir for 
an active heat engine. This passive heat $Q_h$ only accounts for the contributions coming from the energy changes of the 
system associated with transitions between the two states during the part of the period for which the inverse temperature is $\beta_h$. This $Q_h$ does not account for 
dissipated heat associated with ATP consumption.

The first solution for the issue $\eta_{\textrm{ps}}>\eta_C$ is to account for the average entropy production associated with the consumption of ATP during a cycle $\Delta S_{\textrm{act}}$. 
The standard second law from stochastic thermodynamics for this model reads
\begin{equation}
\Delta S_{\textrm{act}}+(\beta_c-\beta_h)Q_h-\beta_cW\ge 0.
\label{eqseclaw2}
\end{equation}
The condition $\eta_{\textrm{ps}}>\eta_C$ is then fully consistent with this second law since $\eta_{\textrm{ps}}$ can be larger than $\eta_C$ if $\Delta S_{\textrm{act}}>0$.

\begin{figure}
\includegraphics[width=8cm]{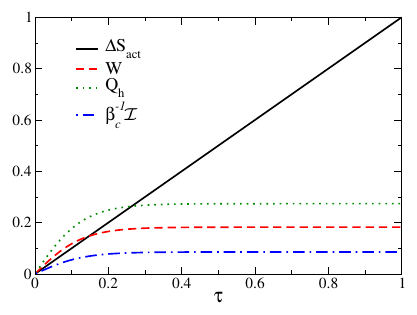}
\vspace{-2mm}
\caption{The different scaling with $\tau$ of  $\Delta S_{\textrm{act}}$ in comparison to $W$, $Q_h$ and $\hh$ for an active 
heat engine with fixed $\Delta\mu$. The inequality $\Delta S_{\textrm{act}}\ge \hh$ holds true for any value of $\tau$. 
The parameters are set to $\beta_c=1$, $\beta_h=0.1$, $E=0.5$, $\delta E=1$, $k=10$, and $\Delta \mu=2$.
}   
\label{fig3} 
\end{figure}

However, it is often the case that this second law does not produce a relevant bound on the rate of extracted work $W$. 
Let us consider the period of the engine $\tau$. For large $\tau$, the amount of entropy produced due to the ATP consumption $\Delta S_{\textrm{act}}$ increases linearly with $\tau$.
This  scaling with the period $\tau$ is different from the scaling of $Q_h$ and $W$. For instance, 
the stochastic work that can be extracted in a cycle is at most $\delta E$. Hence, for large $\tau$ the average work extracted per cycle cannot be more than the constant $\delta E$.
This difference in scaling is illustrated in Fig. \ref{fig3}. We conclude that in the regime of large $\tau$ the standard second law in Eq. \eqref{eqseclaw2} 
does not produce a relevant bound, $\Delta S_{\textrm{act}}$ dominates the inequality, which simply  states that $\Delta S_{\textrm{act}}$ is positive.
 
What is the appropriate statement of the second law that does not suffer from the limitation discussed above? Before we answer this question 
let us consider the following question. How can an active heat engine deliver more work than a passive heat engine? If we introduce a non-zero $\Delta\mu$ in our model the probability of a state is 
shifted. If this shift favors work extraction, then the active heat engine will deliver more work than its passive counterpart.

Our main finding is that the appropriate statement of the second law for active heat engines come from a quantity known as excess entropy. A main interest in this  quantity in stochastic thermodynamics  
comes from the fact that it fulfills a fluctuation theorem \cite{seif12}. Here we show that it has a central role for cyclic active heat engines. For the present model 
this second law derived from the excess entropy reads 
\begin{equation}
\hh+(\beta_c-\beta_h)Q_h-\beta_cW\ge 0.
\label{eqseclaw3}
\end{equation}
The term $\hh$ is the novel term compared to Eq. \eqref{eqseclaw2}. It has an information theoretic expression and it quantifies the shift in the probability of a state due to a 
non-zero $\Delta\mu$. A positive $\hh$ allows an active heat engine to operate beyond the limits that are valid for a passive heat engine, as illustrated 
in Fig. \ref{fig2}. 

Concerning the scaling with the period $\tau$, $\hh$ scales in the same way as $W$ and $Q_h$. Furthermore, we show that $\Delta S_{\textrm{act}}\ge \hh$ independently  
of the value of $\tau$, as illustrated in Fig. \ref{fig3}. The second law above is consistent with $\eta_{\textrm{ps}}>\eta_C$. For a passive heat engine $\hh$ is zero and Eq. \eqref{eqseclaw3} 
becomes the known  second law for passive heat engines in Eq. \eqref{eqseclaw1}.
 
This minimal two-state model does not allow for the discussion of the issue of coarse-graining. For the case with possible hidden dissipative degrees 
of freedom we obtain a second law that has the exact same structure of Eq. \eqref{eqseclaw3}, as shown in Sec. \ref{sec4}.

\section{General Framework}
\label{sec3}

\subsection{Invariant distribution and accompanying distribution}

We consider Markov processes with discrete states. 
A transition rate from state $i$ to state $j$ is denoted $k_{ij}(t)$. Since we want to model cyclic engines, the transition rate is 
periodic in time with a period $\tau$, i.e.,  $k_{ij}(t)= k_{ij}(t+\tau)$. The master equation for the probability to be in state $i$ at time $t$ is 
\begin{equation}
\frac{d}{dt}P_i(t)=\sum_{j}\left[P_j(t) k_{ji}(t)-P_i(t) k_{ij}(t)\right].
\label{eqME}
\end{equation}
The following condition is assumed to hold, if $k_{ij}(t)\neq 0$ then $k_{ji}(t)\neq 0$. In the long time limit, $P_i(t)$ becomes an invariant time-periodic
 distribution with period $\tau$. We denote this invariant distribution simply by $P_i(t)$.

The accompanying distribution \cite{hang82,chet11} $P_i^S(t)$ is generally different from $P_i(t)$. It is defined as the stationary distribution 
the system would reach if the transition rates were fixed and given by $k_{ij}(t)$. Hence, $P_i^S(t)$ is also time-periodic with period $\tau$. 
Specifically, $P_i^S(t)$ is the solution of the equation   
\begin{equation}
\sum_{j}\left[P^S_j(t) k_{ji}(t)-P^S_i(t) k_{ij}(t)\right]=0,
\label{eqProb}
\end{equation}
which is  the stationary form of Eq. \eqref{eqME} for fixed $t$.
If each term in the sum in $j$ is zero, i.e., if detailed balance is fulfilled,  then this accompanying 
distribution $P^S_i(t)$ is an equilibrium distribution denoted by $P^{eq}_i(t)$. Otherwise, the accompanying  distribution  
corresponds to a non-equilibrium stationary distribution denoted by $P_i^S(t)$.

We reiterate the difference between three probability distributions we consider here. 
First, $P_i(t)$ is the time-periodic distribution of the system, corresponding to the long time 
solution of Eq. \eqref{eqME}. Second, $P^S_i(t)$ is the accompanying distribution  corresponding to a 
nonequilibrium steady state. Third, $P^{eq}_i(t)$ is the accompanying distribution corresponding to an equilibrium steady state.

\subsection{Generalized detailed balance}

The periodicity of the transition rates comes from the fact that energies and temperature are periodic. The  energy of state $i$ at time $t$ is denoted $E_i(t)=E_i(t+\tau)$. 
The inverse temperature remains between $\beta_h$ and $\beta_c$ and is written as 
\begin{equation}
\beta(t)= \beta_c(1-\eta_C h(t)),
\label{eqbeta}
\end{equation}
where $h(t)$ is a dimensionless time-periodic function bounded by the inequalities $0\le h(t)\le 1$. If $h(t)=0$ then $\beta(t)=\beta_c$ and if $h(t)=1$ then $\beta(t)=\beta_h$.

An active heat engine must contain internal thermodynamic forces or affinities. An internal affinity is denoted  by $\A_\alpha$. 
We use the index $\alpha$ since the system can have more than one internal affinity. The generalized distance associated with this affinity for a transition from $i$ to $j$ is 
$d_{ij}^{\alpha}(t)=d_{ij}^{\alpha}(t+\tau)$. This distance is antisymmetric, i.e., $d_{ij}^{\alpha}(t)=-d_{ji}^{\alpha}(t)$. 
For example, if $\A_\alpha$ is a force applied to a colloid then $d_{ij}^{\alpha}(t)$ is the spatial distance between $i$ and $j$.

The generalized detailed balance relation \cite{seif12} is a postulate of stochastic thermodynamics that relates  transition rates to physical parameters such as temperature, free energy, and affinities.
This relation is written in the form 
\begin{equation}
\ln\frac{k_{ij}(t)}{k_{ji}(t)}= \beta(t)\left[\Delta E_{ji}(t)+(\beta_c)^{-1}\sum_\alpha\A_\alpha d_{ij}^{\alpha}(t)\right],
\label{eqgendb}
\end{equation}
where $\Delta E_{ji}(t)\equiv E_i(t)-E_j(t)$ and the product $\A_\alpha d_{ij}^\alpha(t)$ has dimension of energy in units of $\beta_c^{-1}$. 
There is no loss of generality in assuming a constant $\A_\alpha$ since a time-dependence of $\A_\alpha$ can be absorbed by the time-dependence of the distance $d_{ij}^\alpha(t)$.

We can now give a clear mathematical definition for active and passive heat engines. If $\A_\alpha=0$ for all $\alpha$ then the heat engine is a passive heat engine. In this case 
the stationary distribution the system would have if the protocol was frozen, i.e., the accompanying distribution,  which corresponds to the solution of Eq. \eqref{eqProb}, 
is the Boltzmann distribution $P^{eq}_i(t)= \textrm{e}^{-\beta(t)E_i(t)}/Z(t)$, where  $Z(t)=\sum_i\textrm{e}^{-\beta(t)E_i(t)}$ is the partition function. Note that a passive 
heat engine is a nonequilibrium system due to the periodic variation of temperature and energies. For nonzero  $\A_\alpha$  
the accompanying distribution $P^{S}_i(t)$ is not the equilibrium distribution and the heat engine is active.


\subsection{Entropy production}

The elementary (probability) current from state $i$ to state $j$ is defined as 
\begin{equation}
J_{ij}(t)\equiv P_i(t) k_{ij}(t)-P_j(t) k_{ji}(t).
\label{eqelem}
\end{equation}
The mathematical form of the average entropy production during a period is \cite{seif12}
\begin{equation}
\Delta S\equiv  \int_0^\tau dt \sum_{i<j} J_{ij}(t)\ln\frac{k_{ij}(t)}{k_{ji}(t)}\ge 0,
\label{eqent}
\end{equation}
where the sum $\sum_{i<j}$ is over all pairs of states. Using the generalized detailed balance relation in Eq. \eqref{eqgendb} we obtain 
\begin{align}
\Delta S=  \sum_\alpha\A_\alpha J_\alpha+ \int_0^\tau dt\sum_{i<j} J_{ij}(t)\beta(t)\Delta E_{ji}(t),
\label{eqent2} 
\end{align}
where
\begin{equation}
J_\alpha\equiv   \int_0^\tau dt\sum_{i<j} (\beta_c)^{-1}\beta(t)J_{ij}(t)d_{ij}^{\alpha}(t). 
\label{eqjalpha}
\end{equation}

The second term in the expression for $\Delta S$ in Eq. \eqref{eqent2}  can be written as
\begin{align}
 \int_0^\tau dtJ_{ij}(t)\beta(t)\Delta E_{ji}(t)=  \int_0^\tau dt \beta_c J_{ij}(t) \Delta E_{ji}(t)+ \eta_CJ_q,
\label{eqsecondterm}
\end{align}
where 
\begin{equation}
J_q\equiv \int_0^\tau dt\sum_{i<j}  h(t)\beta_cJ_{ij}(t) \Delta E_{ij}(t).
\label{eqjq} 
\end{equation}
The physical interpretation for this term is as follows. The energy change of the system in a jump from $i$ to $j$ is $\Delta E_{ij}(t)$. 
The probability current  $J_{ij}(t)$ is the average number of jumps from $i$ to $j$ minus the average number of jumps from $j$ to $i$ per unit of time. 
By summing over all pair of states we have the average energy gain of the system at time $t$. The multiplication by $\beta_c$ means 
that energies are measured in  units of $\beta_c^{-1}$. If the engine is passive the term $J_q$ is a generalized heat flux during a period, since the energy change in a jump is the 
energy taken from the thermal reservoir.  
 
The generalized heat flux  $J_q$ for an active heat engine is not the sole contribution to heat flux 
since the internal affinities $\A_\alpha$ can also contribute to heat dissipation. However, as we show next there 
is a refined first law that also holds true for an active heat engine and does not involve heat dissipation due to the 
internal affinities $\A_\alpha$. We refer to $J_q$ as  passive generalized heat flux.

For the models analyzed here, the inverse temperatures only takes the two values $\beta_c$ and $\beta_h$ during the period $\tau$.  
The function $h(t)$ only takes the values $h(t)= 0$ for $\beta(t)=\beta_c$ and $h(t)= 1$ for $\beta(t)=\beta_h$. For this case, 
$J_q=\beta_c Q_h$, where $Q_h$ is the heat taken from the hot reservoir during a period.

We point out that even though $J_q$ is not the only contribution to heat this quantity can be measured in an experiment. One way to measure it 
would be to observe the trajectory of the system and sum up $h(t)\Delta E_{ij}(t)$ whenever there is a jump from $i$ to $j$. 
An alternative way to measure it would be to evaluate the total heat flux and the heat flux due to the 
activity. Hypothetically, the first quantity could be measured by monitoring small temperature differences in a large but finite reservoir and the second one 
by monitoring changes in the concentration of a chemical fuel such as ATP. The difference between both gives the passive contribution.

\subsection{Refined first law and standard second law}

The entropy production in Eq. \eqref{eqent2} can be written as 
\begin{align}
\Delta S=   \sum_\alpha\A_\alpha J_\alpha+\eta_C J_q - \int_0^\tau dt\sum_{i<j} J_{ij}(t)\beta_c\Delta E_{ij}(t),
\label{eqent3} 
\end{align}
where we used Eq. \eqref{eqsecondterm}. The time-derivative of the average energy $\overline{E}(t)=\sum_iP_i(t)E_i(t)$ reads
\begin{equation}
\frac{d}{dt} \overline{E}(t)= \sum_{i<j} J_{ij}(t)\Delta E_{ij}(t)+\sum_iP_i(t)\frac{d}{dt}E_i(t).
\end{equation} 
Since $\overline{E}(t)=\overline{E}(t+\tau)$, integration over a period leads to 
\begin{equation}
\int_0^\tau dt\sum_{i<j} J_{ij}(t)\Delta E_{ij}(t)=-\int_0^\tau dt\sum_iP_i(t)\frac{d}{dt}E_i(t)\equiv W.
\label{eqW}
\end{equation} 
The term on the right hand is the extracted  work $W$ from stochastic thermodynamics \cite{seif12}. This equation constitutes 
the refined first law for active heat engines. If the inverse temperature only take the values $\beta_c$ and $\beta_h$, i.e., $h(t)$ is either $0$
or $1$,  this equation becomes 
\begin{equation}
Q_h+(-Q_c)= W,
\end{equation} 
where 
\begin{equation} 
-Q_c=-\int_0^\tau dt[1-h(t)]\sum_{i<j} J_{ij}(t)\Delta E_{ij}(t)
\end{equation} 
is the average passive heat delivered to the cold reservoir during a period. We reiterate that this refined first law for active 
heat engines does not involve energy dissipation associated with the affinities $\A_\alpha$.

The entropy production  in Eq. \eqref{eqent3} then becomes
\begin{align}
\Delta S=   \Delta S_{\textrm{act}}+\eta_C J_q -\beta_c W\ge 0,
\label{eqent5} 
\end{align}
where  $\Delta S_{\textrm{act}}\equiv\sum_\alpha\A_\alpha J_\alpha$ is the contribution due to the internal affinities. 
This contribution to the entropy production is the one that makes the heat engine active. This standard form of the second law
has been obtained in \cite{ray17}.

A key issue is the scaling with the period $\tau$ of the different terms in this entropy production. 
A cycle in the network of states is a sequence of jumps that finishes in the state it started. Consider 
a cycle that happens while the external protocol is fixed. The total change of energy associated with this cycle, i.e., the net exchange of passive heat,  is zero.  
The entropy change during the cycle  associated with $\A_\alpha$ 
is not necessarily zero since the distance $d_{ij}^\alpha(t)$ cannot be written as a difference such as $\Delta E_{ij}(t)=E_{j}(t)-E_{i}(t)$. 
The larger period $\tau$, the larger the number of times cycles are performed. Therefore, the active part of the entropy change increases indefinitely 
with $\tau$, unlike passive heat and work.  

This difference in scaling is an important limitation of the second law in Eq. \eqref{eqent5}. For large enough $\tau$ this second law does 
not provide a relevant bound on $W$ since the term  $\Delta S_{\textrm{act}}$ dominates the inequality. A relevant bound on extracted work can only 
contain terms that scale with $\tau$ is the same way that $W$ does. We point out that it is possible to make $\Delta S_{\textrm{act}}$ scale in the 
same way as heat and work with $\tau$ by using infinite energy barriers that blocks the system from performing cycles in the space of states during the period. 
Models for molecular pumps do have  such energy barriers \cite{raha11,ray17}.

\subsection{Excess entropy and second law for active heat engines}

The entropy change $\Delta S$ can be written as  
\begin{equation}
\Delta S= \Delta S_{\textrm{ex}}+\Delta S_{\textrm{hk}},
\end{equation}
where $\Delta S_{\textrm{ex}}$ is the excess entropy and $\Delta S_{\textrm{hk}}$ is the house keeping entropy. 
A main interest in these quantities within stochastic thermodynamics  is related to the fact that each of these terms is positive and, their fluctuating 
versions, each fulfill a separate fluctuation theorem \cite{seif12}. Here we are interested in the mathematical expression of the excess entropy 
for periodically driven systems. As shown in the appendix, the average excess entropy change during a period is given by
\begin{equation}
\Delta S_{\textrm{ex}}= \int_0^\tau dt \sum_{i<j} J_{ij}(t)\ln\frac{P^S_j(t)}{P^S_i(t)}\ge 0.
\label{eqexc}
\end{equation}
We now show that this inequality provides the appropriate statement of the second law for active heat engines.

The main step is to define
\begin{equation}
\hh\equiv  \int_0^\tau dt \sum_{i<j} J_{ij}(t)\left[\ln\frac{P^S_j(t)}{P^{eq}_j(t)}-\ln\frac{P^S_i(t)}{P^{eq}_i(t)}\right].
 \label{eqhh}
\end{equation}
The excess entropy in Eq. \eqref{eqexc} can be written as 
\begin{equation}
\Delta S_{\textrm{ex}}= \hh+\int_0^\tau dt \sum_{i<j} J_{ij}(t)\ln\frac{P^{eq}_j(t)}{P^{eq}_i(t)}.
\label{eqeqexc2}
\end{equation}
The relation  $\ln\frac{P^{eq}_j(t)}{P^{eq}_i(t)}=\beta(t)\Delta E_{ji}(t)$ together with Eq. \eqref{eqjq} and Eq. \eqref{eqW}, leads to
our second law for active heat engines 
\begin{equation}
\Delta S_{\textrm{ex}}= \hh+  \eta_C J_q-\beta_cW\ge 0.
\label{eqeqexc3}
\end{equation}
This inequality involves the extracted work $W$, the generalized heat flux $J_q$, and the new term $\hh$. 

Our second law allows for the following insight about active heat engines. For an active 
heat engine there is a shift from $P^{eq}$ to $P^{S}$, this shift can allow for a larger amount of work extraction, beyond the bound 
given by the second law for a passive heat engine. The amount of energy dissipated by the active process is not directly relevant,
rather the way the probability $P^{S}$ is shifted in relation to $P^{eq}$, as quantified by $\hh$,  is the relevant limiting factor on $W$.

For an active heat engine the pseudo-efficiency $\eta_{\textrm{ps}}$ can go beyond the Carnot bound $\eta_C$.
If we consider the efficiency
\begin{equation}
\eta\equiv  \frac{W}{\beta_c^{-1}J_q+\hh/(\beta_c-\beta_h)},
\label{eqeff}
\end{equation}
which accounts for the term $\hh$, then the second law for active heat engines in Eq. \eqref{eqeqexc3} becomes the traditional 
statement of the second law $\eta\le \eta_C$. We note that an active heat engine is able to extract work even if the inverse temperature is 
fixed and equal to $\beta_c$, as we show with specific examples in Sec. \ref{sec5}. In this case, $W\le \beta_c^{-1}\hh$.

Our second law in Eq. \eqref{eqeqexc3} has the correct scaling with the period $\tau$ and becomes the second law for passive heat engines for the 
case $P^S=P^{eq}$ during the whole period. First, for a cycle in the space of states the change of $\ln\frac{P^S_i(t)}{P^{eq}_i(t)}-\ln\frac{P^S_j(t)}{P^{eq}_j(t)}$ is zero. Hence, 
for large $\tau$, the scaling with $\tau$ of the term $\hh$ is the same as the scaling of $W$ and $J_q$. Second, for a passive heat engine $\hh=0$ and Eq.\eqref{eqeqexc3} becomes the 
standard second law for passive heat engines.

From the second law for the housekeeping entropy \cite{seif12}, we obtain 
\begin{equation}
\Delta S_{\textrm{hk}}= \Delta S- \Delta S_{\textrm{ex}}\ge 0.
\label{eqhk}
\end{equation}
This inequality, together with Eq. \eqref{eqent5} and Eq. \eqref{eqeqexc3}, leads to
\begin{equation}
\Delta S_{\textrm{act}}\ge \hh.
\label{eqsh}
\end{equation}
Hence, the second law in Eq. \eqref{eqeqexc3} provides a tighter bound on the extracted work $W$ than the one in Eq. \eqref{eqent5},  for any period $\tau$.
The term $\hh$ is the information theoretic contribution to the second law for an active heat engine and $\Delta S_{\textrm{act}}$ can be viewed as 
the total thermodynamic cost to "generate" this term. This total thermodynamic cost to generate an information-theoretic term shows up in many examples 
in the relation between information and thermodynamics \cite{benn82,deff13,hart14,horo14}.

We here assumed that there is only one transition path between a pair of states $i,j$. However, it is possible to have more than one transition path between a pair of states, 
which is the case for the two-sate model in Sec. \ref{sec52}. The generalization of the expressions obtained here for the case that includes this possibility of more than 
one link between a pair of states is straightforward. If we use the label $\nu$ to represent the different links, the transition rate
from $i$ to $j$ through link $\nu$ is $k_{ij}^{\nu}$. For the generalized detailed balance relation in Eq. \eqref{eqgendb}, the distances  $d_{ij}^{\alpha}(t)$ depend on the index $\nu$. 
The expression for quantities that show up in the second law inequalities remain similar, with the sum $\sum_{i<j}$ substituted by $\sum_\nu\sum_{i<j}$. 

The second law obtained here is relevant for generic periodically driven active heat engines. Another important class of heat engines are those driven by constant time-independent
forces that operate in a steady state. Since the average excess entropy in Eq. \eqref{eqexc} is zero in a steady state, it provides no relevant information about the performance 
of steady state heat engines, which can be quantified with the standard entropy production \cite{piet19}.   

\subsection{Information theoretic expression for $\hh$}

It turns out that $\hh$ can be expressed as part of a time derivative of a difference of Kullback-Leibler distances. Consider the average 
\begin{equation}
\langle\ln \frac{P^S}{P^{eq}}\rangle(t)\equiv \sum_iP_i(t)\ln\frac{P^S_i(t)}{P^{eq}_i(t)}. 
\end{equation}
This term can be written as 
\begin{equation}
\langle\ln \frac{P^S}{P^{eq}}\rangle(t)= D_{KL}[P||P^{eq}](t)-D_{KL}[P||P^{S}](t),
\label{eqdkl}
\end{equation}
$D_{KL}[p||q]\equiv \sum_ip_i\ln(p_i/q_i)$ is the Kullback-Leibler distance.

Taking a a derivative of this term we obtain
\begin{align}
\frac{d}{dt}\langle\ln \frac{P^S}{P^{eq}}\rangle= &  \sum_{i<j}J_{ij}(t)\left[\ln\frac{P^S_j(t)}{P^{eq}_j(t)}-\ln\frac{P^S_i(t)}{P^{eq}_i(t)}\right]\nonumber\\
&+\sum_iP_i(t)\frac{d}{dt}\left(\ln\frac{P^S_i(t)}{P^{eq}_i(t)}\right).
\end{align}
Since $\langle\ln \frac{P^S}{P^{eq}}\rangle(t)=\langle\ln \frac{P^S}{P^{eq}}\rangle(t+\tau)$, integration of the equation above yields 
\begin{align}
& \int_0^\tau dt \sum_{i<j}J_{ij}(t)\left[\ln\frac{P^S_j(t)}{P^{eq}_j(t)}-\ln\frac{P^S_i(t)}{P^{eq}_i(t)}\right]=\nonumber\\
& -\int_0^\tau dt \sum_iP_i(t)\frac{d}{dt}\left(\ln\frac{P^S_i(t)}{P^{eq}_i(t)}\right).
\end{align}
This equality provides a different expression for $\hh$ in Eq. \eqref{eqhh}, which is 
\begin{equation}
\hh=-\int_0^\tau dt \sum_iP_i(t)\frac{d}{dt}\left(\ln\frac{P^S_i(t)}{P^{eq}_i(t)}\right).
\label{eqhh3}
\end{equation}
Hence, the term $\hh$ is part of the time-derivative of the difference of Kullback-Leibler distances in Eq. \eqref{eqdkl}. 
Furthermore, the expression in Eq. \eqref{eqhh3} can be more convenient to calculate $\hh$. 

\section{Coarse-grained second law}
\label{sec4}

\subsection{Observable variable}

We now address the issue of coarse-graining. The first step is to divide the state $i$ into two variables $i=(x,a)$. The variable $x$ 
can be observed and the variable $a$ is hidden. If we consider the experiment from Ref. \cite{kris16}, $x$  would be the position of the colloidal particle 
and $a$ would be related to the degrees of freedom that determine the state of the bacteria. If we consider 
the variable $a$ as part of the reservoir then we can say that the system labeled by the variable $x$ is in contact with a nonequilibrium or active reservoir.  

The master equation in Eq. \eqref{eqME} implies an 
equation for the time evolution of  $P_x(t)\equiv\sum_aP_{x,a}(t)$. Writing $k_{ij}$ as $k_{a,x;a',x'}$ and with the definition $K_{xx'}(t)=\sum_{a,a'}P_{a|x}(t) k_{x,a;x',a'}(t)$, 
where $P_{a|x}(t)\equiv P_{x,a}(t)/P_x(t)$ is a conditional probability, this equation becomes 
\begin{equation}
\frac{d}{dt}P_x(t)=\sum_{x'}\left[P_x(t) K_{xx'}(t)-P_{x'}(t) K_{x'x}(t)\right].
\label{eqMEx}
\end{equation}
We point out that the dynamics of the variable $x$ is non-Markovian since  $K_{xx'}$ depends on the dynamics of the hidden degree of freedom represented by $a$.

The transition rates $k_{x,a;x',a'}(t)$ fulfill the generalized detailed balance relation from Eq. \eqref{eqgendb}. The affinities $\A_\alpha$ are related to the variable 
$a$ and they are responsible for the active nature of the engine. The accompanying distribution  is still 
the solution of Eq. \eqref{eqProb}. For the coarse-grained variable $x$ we have $P_x^S=\sum_aP_{x,a}^S$.

\subsection{Passive counterpart}

We assume that the time-dependence of the energy $E_i(t)=E_{x,a}(t)$ is contained only in the observable variable $x$.
If this assumption does not hold we would also need access to the variable $a$ to evaluate the work.
This energy can then be written as  
\begin{equation}
E_{x,a}(t)= E_{x}(t)+ E^{\textrm{int}}_{x,a}+ E_a, 
\end{equation}
where $E^{\textrm{int}}_{x,a}$ represents the interaction energy between $x$ and $a$. 
 
The passive counterpart for our active heat engine is defined as follows. The equilibrium accompanying distribution associated with this passive counterpart 
is
\begin{equation}
P_{x}^{eq}(t)= \textrm{e}^{-\beta E_x(t)}/Z(t),
\label{eqpeq}
\end{equation}
where $Z(t)= \sum_x\textrm{e}^{-\beta E_x(t)}$. Such a passive heat engine corresponds to the complete absence of the degrees of freedom represented by the variable 
$a$. Physically, if we consider the experiment in \cite{kris16}, this passive counterpart corresponds to an experiment where all bacteria are removed from 
the solution. 

A different passive counterpart would be to keep the variable $a$ but set all affinities $\A_\alpha$ to zero. In this case, the accompanying equilibrium distribution for the passive counterpart would be 
\begin{equation}
p_{x}^{eq}(t)= \sum_a P_{x,a}^{eq}\propto \textrm{e}^{-\beta(t) E^{\textrm{eff}}_x(t)}.
\label{eqpeqalt}
\end{equation}
where $E^{\textrm{eff}}_x(t)= E_x(t)-\beta(t)^{-1}\ln\left[\sum_a\textrm{e}^{-\beta(t)(E^{\textrm{int}}_{x,a}+ E_a})\right]$. The difference is that this effective energy 
also contains the interaction energy with the degrees of freedom represented by $a$. If we consider the experiment in \cite{kris16}, this would correspond
to an experiment with all bacteria present but with their internal driving forces set to zero. 

We proceed with the elimination of the variable $a$ as the passive counterpart of the engine. However, this choice does not represent a limitation in our framework.
As we show below, our formalism also applies if we choose this second type of passive counterpart of the engine, with modified expressions for the terms in the second law for active heat engines.

\subsection{Work, heat and entropy}

The work in Eq. \eqref{eqW} can be written as    
\begin{equation}
W= -\int_0^\tau dt\sum_{x}P_{x}(t)\frac{d}{dt}E_{x}(t).
\label{eqW2}
\end{equation}
This expression follows from the fact that the time-dependence of $E_{x,a}(t)$ is fully contained in $E_{x}(t)$.
The coarse-grained generalized heat flux is defined as 
\begin{equation}
J_q^{\textrm{cg}}\equiv \int_0^\tau dt\sum_{x<x'}  h(t)\beta_cJ_{xx'}(t) \Delta E_{xx'}(t),
\label{eqjq2} 
\end{equation}
where $J_{xx'}(t)\equiv P_x(t) K_{xx'}(t)-P_{x'}(t) K_{x'x}(t)$ and $\Delta E_{xx'}(t)\equiv E_{x'}(t)-E_{x}(t)$. This expression is in general different from Eq. \eqref{eqjq}.

The coarse-grained variable $x$ also has a refined first law, which is obtained in the following way. The time-derivative of  the average of the part of the energy that depends on $x$ is
\begin{equation}
\frac{d}{dt}\sum_x P_x(t)E_x(t)= \sum_{x<x'}  J_{xx'}(t) \Delta E_{xx'}(t)+\sum_{x}P_{x}(t)\frac{d}{dt}E_{x}(t).
\label{eqfirstlawx1} 
\end{equation}
From the fact that this average energy is periodic, we obtain 
\begin{equation}
\int_0^\tau dt \sum_{x<x'}  J_{xx'}(t) \Delta E_{xx'}(t)= W, 
\label{eqfirstlawx2} 
\end{equation}
which is the refined first law for the variable $x$.

For the particular case where the inverse temperature only takes two values $\beta_c$ and $\beta_h$ , i.e., $h(t)$ is either $0$ or $1$,  the coarse-grained generalized heat on Eq. \eqref{eqjq2} 
becomes the coarse-grained heat taken from the hot reservoir $Q_h^{\textrm{cg}}$.
The first law in Eq. \eqref{eqfirstlawx2} becomes 
\begin{equation}
Q_h^{\textrm{cg}}+(-Q_c^{\textrm{cg}})= W, 
\label{eqfirstlawx3} 
\end{equation}
where $-Q_c^{\textrm{cg}}\equiv -\int_0^\tau dt\sum_{x<x'}  [1-h(t)]\beta_cJ_{xx'}(t) \Delta E_{xx'}(t)$ is the coarse-grained heat released to the cold reservoir.

The coarse-grained excess entropy, as shown in  the appendix, follows the second law inequality
\begin{equation}
\Delta S_{\textrm{ex}}^{\textrm{cg}}\equiv \int_0^\tau dt \sum_{x<x'} J_{xx'}(t)\ln\frac{P^S_{x'}(t)}{P^S_x(t)}\ge 0.
\label{eqexcg}
\end{equation}
Analogously to the procedure used for  $\Delta S_{\textrm{ex}}$, the coarse-grained excess entropy can be written as 
\begin{equation}
\Delta S_{\textrm{ex}}^{\textrm{cg}}= \hh^{\textrm{cg}}+  \eta_C J_q^{\textrm{cg}}-\beta_cW\ge 0,
\label{eqeqexcg3}
\end{equation}
where
\begin{equation}
\hh^{\textrm{cg}}\equiv \frac{1}{\tau}\int_0^\tau dt \sum_{x<x'} J_{xx'}(t)\left[\ln\frac{P^S_{x'}(t)}{P^{eq}_{x'}(t)}-\ln\frac{P^S_x(t)}{P^{eq}_x(t)}\right]\ge 0.
\label{eqhhcg1}
\end{equation}
Similar to $\hh$, the coarse-grained term $\hh^{\textrm{cg}}$ has the same information theoretical expression as a part of the time-derivative of $\sum_x P_x(t)\ln \frac{P^S_x(t)}{P^{eq}_x(t)}$, 
which leads to the alternative formula for $\hh^{\textrm{cg}}$,
\begin{equation}
\hh^{\textrm{cg}}=-\int_0^\tau dt \sum_xP_x(t)\frac{d}{dt}\left(\ln\frac{P^S_x(t)}{P^{eq}_x(t)}\right).
\label{eqhhcg2}
\end{equation}

Remarkably, the coarse-grained second law in Eq. \eqref{eqeqexcg3} has the exact same form as compared to the full second law in Eq. \eqref{eqeqexc3}, 
in spite of the different values of the coarse-grained quantities. The physical interpretation of $\hh^{\textrm{cg}}$ as an extra resource that allows 
for more extracted work in an active heat engine also applies to this coarse-grained second law.
The standard entropy production of stochastic thermodynamics also has this feature of invariance 
under coarse-graining that we found here for the excess entropy \cite{espo12}. 

The structure of the second law for active heat engines is independent of whether one has full access 
to the state $i$ or only access to $x$. For the first kind of active heat engine the system represented by $i$ is in contact with an equilibrium heat bath and for 
the second kind the system represented by $x$ is in contact with an nonequilibrium heat bath. 

As proved in the appendix the excess entropy is larger than the coarse-grained excess entropy, i.e.,
\begin{equation}
\Delta S_{\textrm{ex}}\ge \Delta S_{\textrm{ex}}^{\textrm{cg}}.
\label{eqexcine}
\end{equation}
Hence, even if we do have access to the hidden variable $a$ the coarse-grained excess entropy provides a tighter bound on extracted work.

The passive counterpart of the engine was so far chosen as the absence of the variable $a$. For a passive counterpart that corresponds to setting all affinities $\A_\alpha$
to zero, the expressions above change as follows. The work $W$ remains the same. The coarse-grained  generalized heat $J_q^{\textrm{cg}}$ has the expression  in Eq. \eqref{eqjq2} 
with $E^{\textrm{eff}}_x(t)$ instead of $E_x(t)$. The term $\hh^{\textrm{cg}}$ has the expression in Eq. \eqref{eqhhcg1} (or the alternative expression in Eq. \eqref{eqhhcg2})
with $p_{x}^{eq}(t)$ in Eq. \eqref{eqpeqalt} instead of $P_{x}^{eq}(t)$ in Eq. \eqref{eqpeq}. The coarse-grained excess entropy $\Delta S_{\textrm{ex}}^{\textrm{cg}}$ is independent 
of the choice of the passive counterpart. 
  
For all the models we consider here, either there are no interactions between $x$ and $a$, i.e., $E^{\textrm{int}}_{x,a}=0$ or,
the condition that $\sum_a \textrm{e}^{-\beta(t)(E^{\textrm{int}}_{x,a}+ E_a)}$ is independent of $x$ holds. 
For these cases $p_{x}^{eq}(t)=P_{x}^{eq}(t)$ and both passive counterparts correspond to the same expression for the terms that show up in 
$\Delta S_{\textrm{ex}}^{\textrm{cg}}$. 
   
In our framework we have considered a master equation with discrete states. Hence, our results are also valid for a continuous state space that 
follows overdamped Langevin dynamics, since an overdamped Langevin equation can be obtained as a limit of a master equation. In fact, we do consider 
two models with continuous variables in the next two sections. The generalization of our framework to underdamped Langevin dynamics is left for future work.

\section{Analytical case studies }
\label{sec5}

\subsection{Expressions for $W$, $Q_h$, $\hh$, and $\hh^{\textrm{cg}}$}
\label{sec51}

\begin{figure}
\includegraphics[width=8cm]{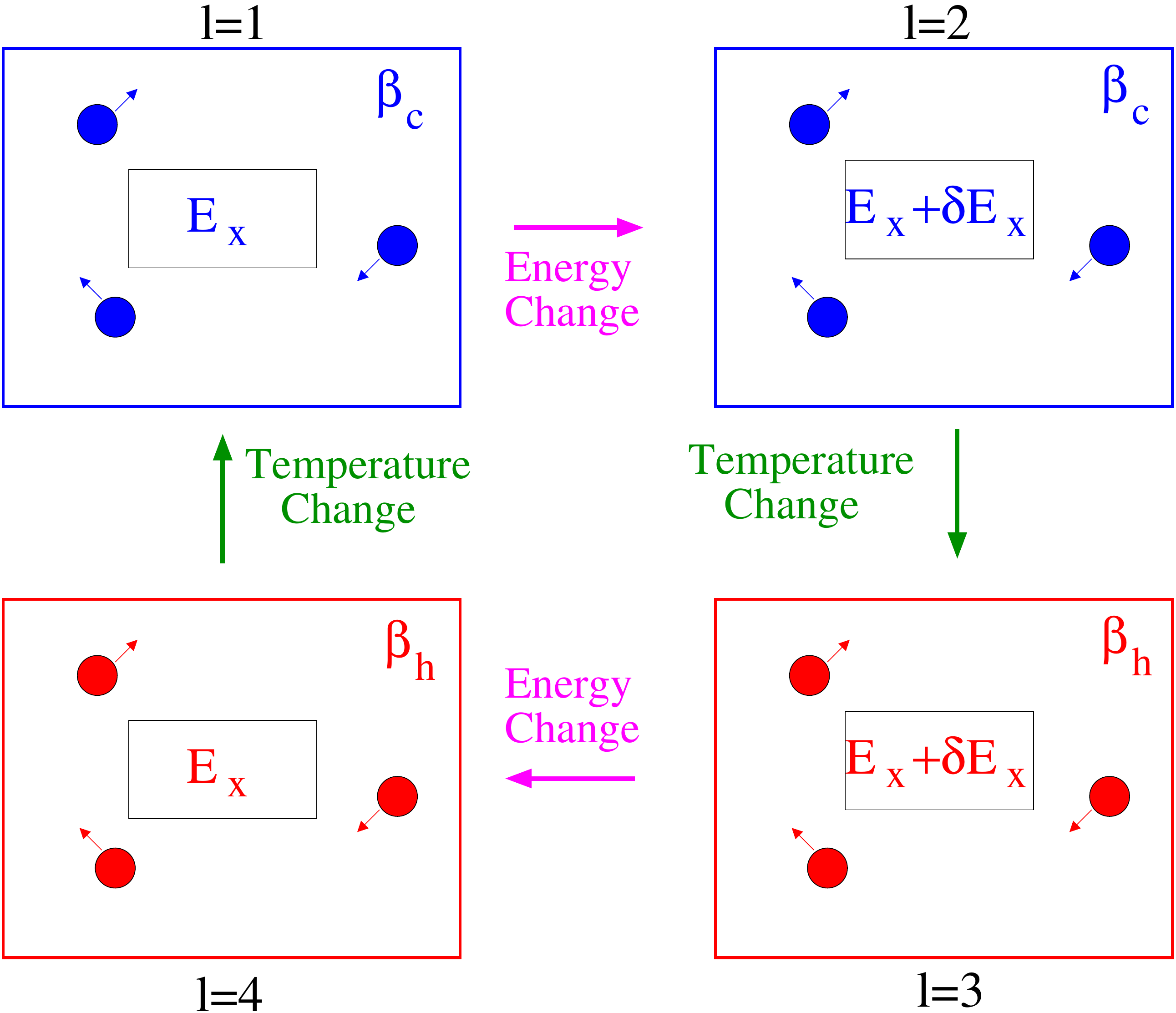}
\vspace{-2mm}
\caption{Depiction of the protocol for the case studies. The inner box represents the observable degree of freedom $x$ and the respective energy $E_x$.
The particles with arrows represent hidden active degrees of freedom labeled by $a$. The four steps are as follows: from $l=1$ to $l=2$ 
energy changes from   $E_x$ to $E_x+\delta E_x$, from $l=2$ to $l=3$ the inverse temperature changes from $\beta_c$ to $\beta_h$, from $l=3$ to $l=4$ the energy changes from 
from $E_x+\delta E_x$ to $E_x$, and from $l=4$ to $l=1$ the inverse temperature changes from $\beta_h$ to $\beta_c$. 
}   
\label{fig4} 
\end{figure}

For the models analyzed in this section the period has four parts illustrated in Fig. \ref{fig4}. Each part has duration $\tau_l$, with $l=1,2,3,4$ and
$\sum_{l=1}^4\tau_l=\tau$. There are two energy changes and two temperature changes. The time-dependent temperature  is $\beta(t)=\beta_c$ for 
$0\le t\le \tau_1+\tau_2$ and $\beta(t)=\beta_h$ for $\tau_1+\tau_2< t\le \tau$. The time-dependent energy is $E_x$ for $0\le t\le \tau_1$
and for $\tau_1+\tau_2+\tau_3\le t\le \tau$, and $E_x+\delta E_x$ for  $\tau_1\le t\le \tau_1+\tau_2+\tau_3$.
Since the accompanying distributions $P^S_x(t)$ and $P^{eq}_{x}(t)$ are also divided into four parts, we write them as $P^{S,l}_x$ and $P^{eq,l}_x$, respectively.  
The affinities $\A_\alpha$ can also change for different parts of the period. 

The times are chosen such that $\tau_1=\tau_3$ and $\tau_2=\tau_4$. Furthermore, the temperature changes are assumed to be instantaneous, which
corresponds to the limit $\tau_2\to 0$. Physically, the temperature changes have to be much shorter than the typical time for a transition between states.
In this limit, $\tau_1=\tau/2$. For the two-state model we do obtain analytical expressions for finite time $\tau$. For all other models we consider 
the limit of large $\tau$, which means that at the end of the first and third parts the system has reached the respective stationary state. This assumption 
simplifies calculations since the distribution $P_x(t)$ at the end of these two parts is the same as the accompanying distribution $P^{S}_x(t)$.
We point out that this case does not correspond to a quasi-static limit since the temperature changes are instantaneous. Within this particular 
limit the expressions for $W$, $Q_h=(\beta_c)^{-1}J_q$, $\hh$, and $\hh^{\textrm{cg}}$ acquire the following forms.

First the work in Eq. \eqref{eqW2} is
\begin{equation}
W= \sum_x (P^{S,3}_x-P^{S,1}_x)\delta E_x.
\label{eqWtimescale}
\end{equation}
This expression can be directly understood from Fig. \ref{fig4} as follows. From the first part to the second part the energy increases by $\delta E_x$, 
which gives the contribution $-P^{S,1}_x\delta E_x$. From the third to the fourth part the energy decreases by  $\delta E_x$, 
which gives the contribution $P^{S,3}_x\delta E_x$.

The heat taken from the hot reservoir in Eq. \eqref{eqjq} is
\begin{equation}
Q_h= \sum_{x,x'} P^{S,1}_xP^{S,3}_{x'}(E_{x'}+\delta E_{x'}-E_x-+\delta E_{x}).
\label{eqQtimescale}
\end{equation}
For the models in this section, the heat flux and the coarse-grained heat flux in Eq. \eqref{eqjq2} are equal. 
The term $\hh$ in Eq. \eqref{eqhh3} becomes
\begin{equation}
\hh= \sum_{x,a} (P^{S,3}_{x,a}-P^{S,1}_{x,a})\left(\ln\frac{P^{S,3}_{x,a}}{P^{eq,3}_{x,a}}-\ln\frac{P^{S,1}_{x,a}}{P^{eq,1}_{x,a}}\right).
\label{eqhtimescale}
\end{equation}
The coarse-grained term $\hh^{\textrm{cg}}$ in Eq \eqref{eqhhcg2} is given by
\begin{equation}
\hh^{\textrm{cg}}= \sum_{x} (P^{S,3}_x-P^{S,1}_x)\left(\ln\frac{P^{S,3}_x}{P^{eq,3}_x}-\ln\frac{P^{S,1}_x}{P^{eq,1}_x}\right).
\label{eqhcgtimescale}
\end{equation}
Since $J_q=J_q^{\textrm{cg}}$ for the models in this section, the inequality in Eq. \eqref{eqexcine} implies $\hh\ge\hh^{\textrm{cg}}$.

\begin{figure*}[t]
\subfigure[]{\includegraphics[width=72mm]{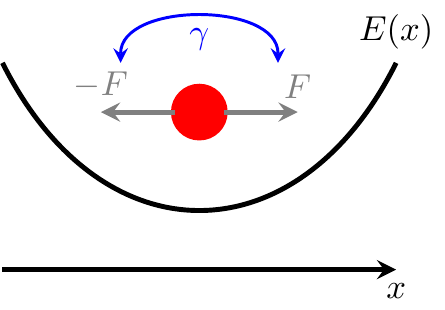}\label{fig5a}}\hspace{3mm}
\subfigure[]{\includegraphics[width=72mm]{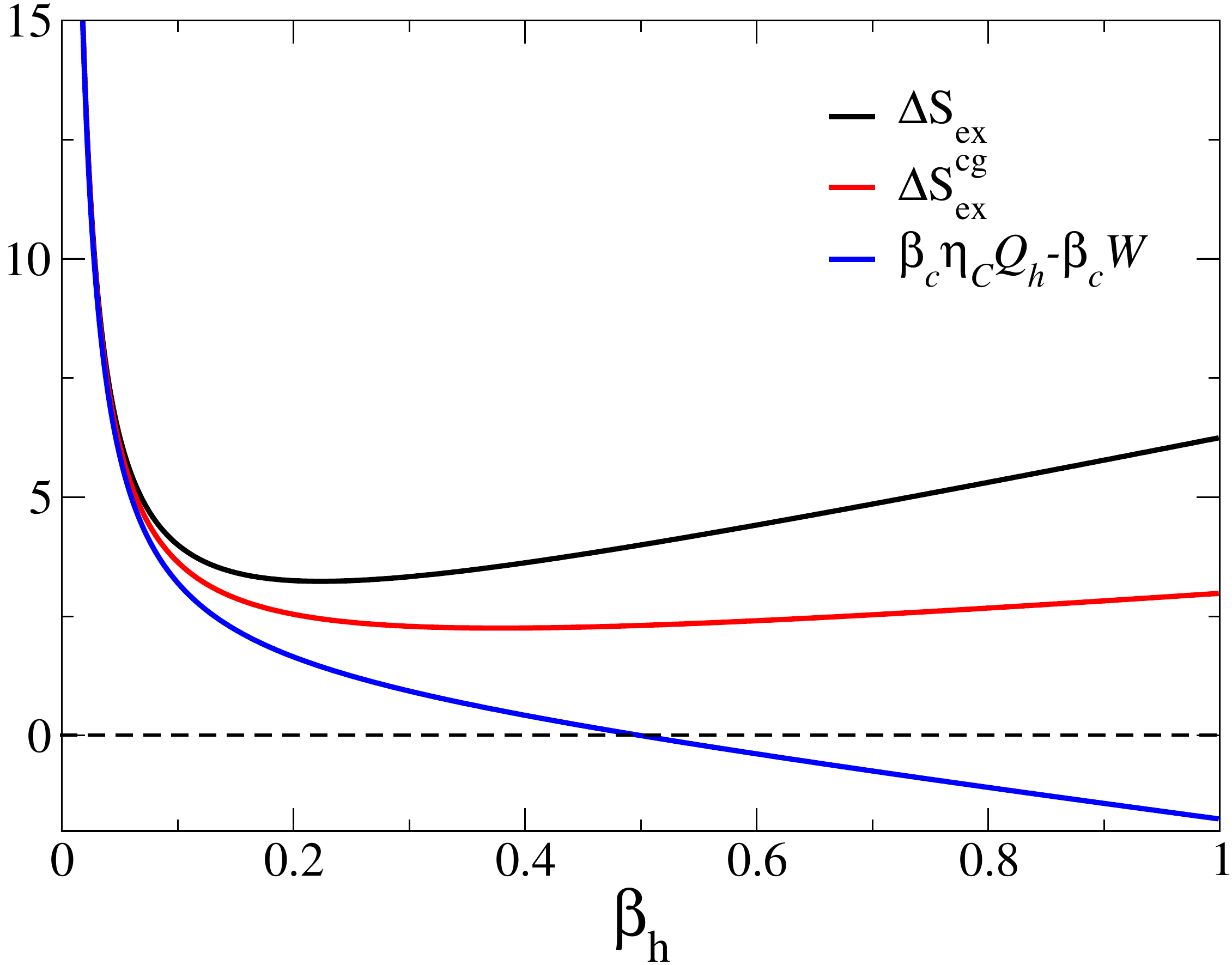}\label{fig5b}}\hspace{3mm}
\subfigure[]{\includegraphics[width=72mm]{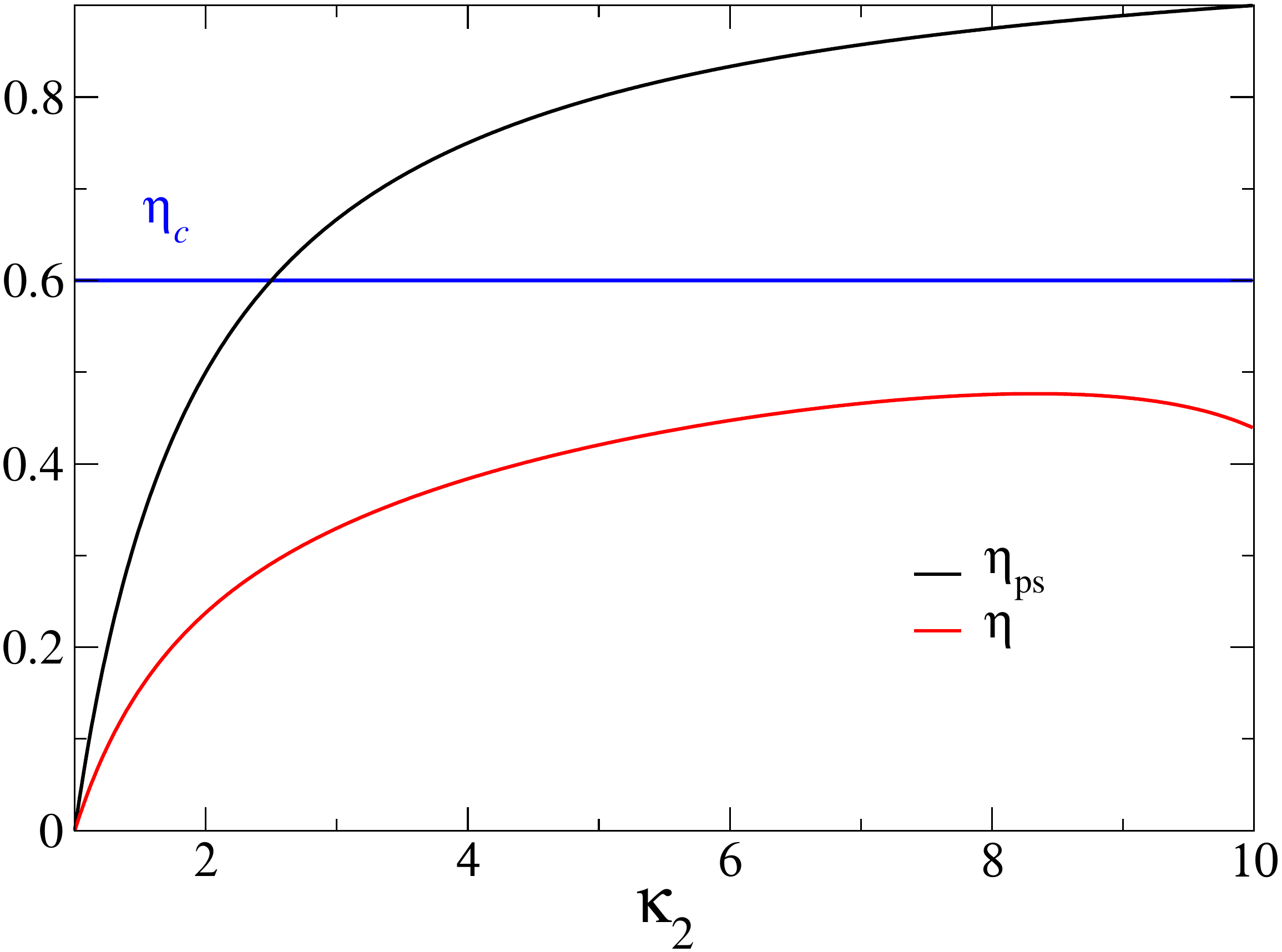}\label{fig5c}}\hspace{3mm}
\subfigure[]{\includegraphics[width=72mm]{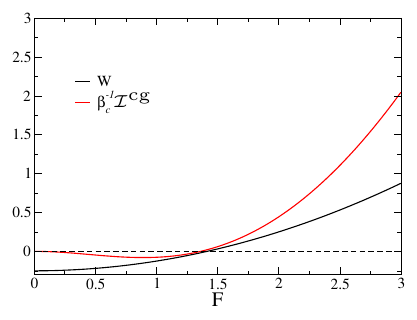}\label{fig5d}}
\vspace{-2mm}
\caption{(a) Depiction of an active particle with a force $F$ that changes sign with a rate $\gamma$ and in
a harmonic potential $E(x)$. (b) The quantities $\beta_c\eta_CQ_h-\beta_cW$, which can be negative for an active heat engine, 
$\Delta S_{\textrm{ex}}^{\textrm{cg}}$, and $\Delta S_{\textrm{ex}}$, as a function of the inverse hot temperature 
$\beta_h$ for $\kappa_2= 2$ and $F=4$. (c) Efficiency $\eta$ and pseudo-efficiency $\eta^\textrm{ps}$ as functions of $\kappa_2$ and compared to the Carnot 
bound for $\beta_h=0.4$, $F=10$. (d) Work $W$ and its upper bound $(\beta_c)^{-1}\hh^{\textrm{cg}}$ as functions of $F$ for constant temperature $\beta_h=\beta_c$
and $\kappa_2=2$. For the three plots $\beta_c=1$ and $\kappa_1=1$.}   
\label{fig5} 
\end{figure*}

\subsection{Two-state model}
\label{sec52}

The two-state model is defined as follows. The transition rates of 
the model are given by $\kappa^+(t)=\omega^-(t)=k$, $\kappa^-(t)=k\textrm{e}^{\beta(t)[E(t)+\Delta\mu]}$, and $\omega^+(t)= k\textrm{e}^{\beta(t)E(t)}$, 
where the rates are represented in Fig. \ref{fig1b}. The time-dependent energy $E(t)$ is $E$ during the first and fourth parts of the period 
and $E+\delta E$ during the second and third parts of the period. The affinity $\A_\alpha$ is the free energy of one ATP hydrolysis $\Delta\mu$. 

Using the methods from Ref. \cite{bara18}, which contains expressions for generic currents in two-state systems, 
we can evaluate $Q_h$, $W$, $\hh$, and $\Delta S_{\textrm{act}}$ analytically for finite $\tau$. The expressions 
are too long to show here. The results for this model are shown in Sec. \ref{sec2}.

\subsection{Active Particle under the action of a harmonic potential}
\label{sec53}

We now consider a model of an active Brownian particle in one dimension under the action of a harmonic potential \cite{dhar19,basu20,garc21}. 
Similar models for cyclic active heat engines have been analyzed in \cite{zaki17,saha19,holu20,kuma20}. Let us first consider the model for 
the case of fixed temperature and energy illustrated in Fig. \ref{fig5a}. The position of the particle is labeled $x$. For this model $x$ is a continuous variable, 
we then write the energy $E_x$ as $E(x)$. The harmonic potential is written as $E(x)= \kappa x^2/2$. This model can be described by an overdamped Langevin  equation that can be obtained as 
the continuous limit of the master equation for discrete $x$.  A force of magnitude $F$ acts on the active Brownian particle. The sign of the force is a Poisson 
process with a rate $\gamma$. The variable $a$ then takes two values, one $a=+1$ corresponding to a force $+F$ and the other $a=-1$ corresponding to 
the force $-F$. The force $F$ multiplied by distance is the affinity $\A_\alpha$ for this model.

We consider the limit for which $\gamma$ is much smaller than the inverse relaxation time of the particle, the particle reaches a stationary state with a given sign for 
the force $F$ before changing the sign of the force. The stationary probability density for the particle to be  at position $x$ with $a=\pm 1$ is  then 
\begin{equation}
P^S_{\pm}(x)= \frac{\sqrt{\beta  \kappa} \textrm{e}^{-\frac{\beta  F^2}{2 \kappa}}\textrm{e}^{-\frac{1}{2} \beta  \kappa x^2\mp\beta  F x}}{2\sqrt{2 \pi }}.
\label{eqPSamodel2}
\end{equation}
Furthermore, the probability density to be at position $x$ is given by      
\begin{equation}
P^S(x)=P^S_{+}(x)+P^S_{-}(x)= \frac{\sqrt{\beta  \kappa} \textrm{e}^{-\frac{\beta  F^2}{2 \kappa}}\textrm{e}^{-\frac{1}{2} \beta  \kappa x^2}}{\sqrt{2 \pi }}\cosh(\beta F x).
\label{eqPSmodel2}
\end{equation}
The equilibrium distribution that corresponds to $F=0$ is given by 
\begin{equation}
P^{eq}(x)= \frac{\sqrt{\beta  \kappa}\textrm{e}^{-\frac{1}{2} \beta  \kappa x^2}}{\sqrt{2 \pi }}.
\label{eqPeqmodel2}
\end{equation}
   
During the first part of the period the stiffness of the harmonic potential is $\kappa_1$ and the force is $F_1$. During the third part of the period the stiffness is $\kappa_2$
and the force is $F_2$. The second and fourth parts of the period are instantaneous. In terms of the notation in Sec. \ref{sec51}, $E_x=\kappa_1x^2/2$ and $E_x+\delta E_x=\kappa_2x^2/2$.

The work $W$ in Eq. \eqref{eqWtimescale}, with an integral over $x$ instead the summation over $x$, is obtained with 
Eq. \eqref{eqPSmodel2}, which leads to the expression    
\begin{equation}
W= \frac{(\kappa_2-\kappa_1) \left[\beta_c
   \kappa_1^2 \left(\beta_h F_2^2+\kappa_2\right)-\beta_h\kappa_2^2 (\beta_cF_1^2+\kappa_1)\right]}{2 \beta_c \beta_h \kappa_1^2 \kappa_2^2}
\end{equation}
The heat taken from the hot reservoir in Eq. \eqref{eqQtimescale} for the present model is
\begin{equation}
Q_h= \frac{\beta_h F_2^2+\kappa_2}{2 \beta_h \kappa_2}-\frac{\kappa_2 \left(\beta_c F_1^2+\kappa_1\right)}{2 \beta_c \kappa_1^2}.
\end{equation}
The term $\hh$ in Eq. \eqref{eqhtimescale} is obtained with Eq. \eqref{eqPeqmodel2} and Eq.\eqref{eqPSamodel2}, which lead to
\begin{equation}
\hh= \frac{(\beta_c F_1-\beta_h F_2) (F_1 \kappa_2-F_2\kappa_1)}{\kappa_1 \kappa_2}
\end{equation}
The coarse-grained term $\hh^{\textrm{cg}}$ in Eq. \eqref{eqhcgtimescale} follows from  Eq. \eqref{eqPSmodel2} and Eq. \eqref{eqPeqmodel2}, 
\begin{align}
\hh^{\textrm{cg}}= & \int_{-\infty}^{\infty}dx\{P^{S,3}(x)\ln[\cosh(F_2\beta_h x)]\nonumber\\
& -P^{S,1}(x)\ln[\cosh(F_1\beta_c x)]\}.
\end{align}
For this quantity the expression is in terms of an integral that has to be performed numerically.

We now restrict to the case $F_1=0$ and $F_2=F$. In Fig. \ref{fig5b} we show that $\beta_c\eta_CQ_h-\beta_cW$ can be negative and that the inclusion of the term $\hh^{\textrm{cg}}$ 
recovers the appropriate statement of the second law for active heat engines. We also illustrate the inequality   $\hh\ge \hh^{\textrm{cg}}$ valid for the models 
analyzed here. In Fig. \ref{fig5c} we illustrate the second law for active heat engines from a different perspective.   We compare the pseudo efficiency $\eta^\textrm{ps}= w/Q_h$ and 
the efficiency $\eta\equiv  \frac{W}{Q_h+\hh^{\textrm{cg}}/(\beta_c-\beta_h)}$ with the Carnot efficiency $\eta_C$. Our second law does become the standard statement
$\eta\le \eta_C$ if we also account for the term   $\hh^{\textrm{cg}}$, while the pseudo-efficiency does cross the Carnot bound. 

This active heat engine can extract work even if the temperature is constant, i.e., $\beta_h=\beta_c$, as shown in Fig. \ref{fig5d}, for which the extracted work 
becomes positive for $F>1.4$ approximately. In this case, the second law becomes $W\le (\beta_c)^{-1}\hh^{\textrm{cg}}$.

\subsection{Two-interacting particles}
\label{sec54}

\begin{figure*}[t]
\subfigure[]{\includegraphics[width=72mm]{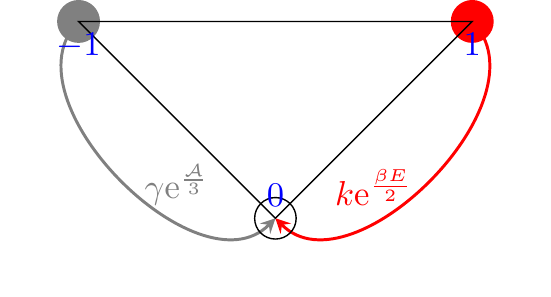}\label{fig6a}}\hspace{3mm}
\subfigure[]{\includegraphics[width=72mm]{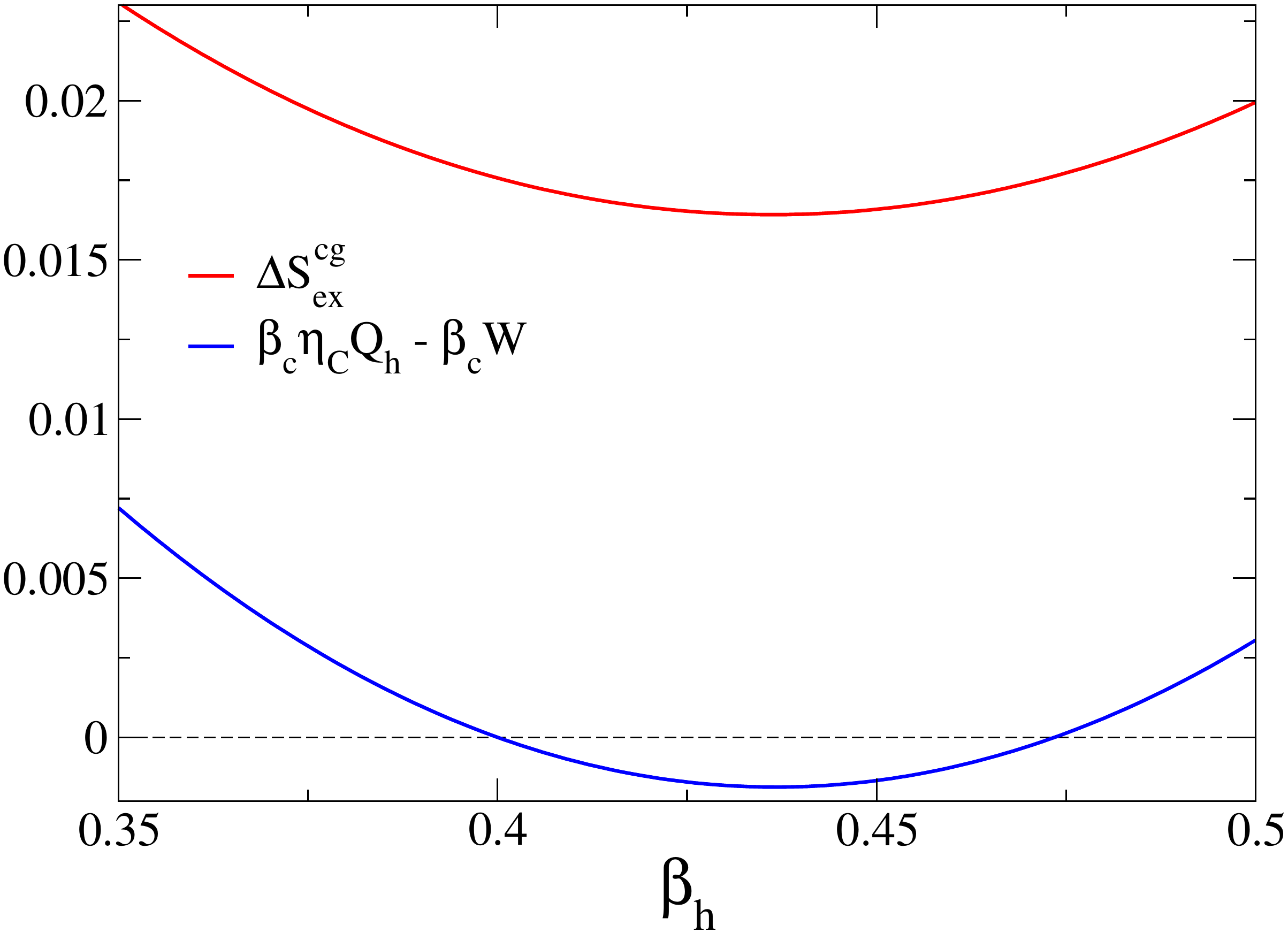}\label{fig6b}}\hspace{3mm}
\subfigure[]{\includegraphics[width=72mm]{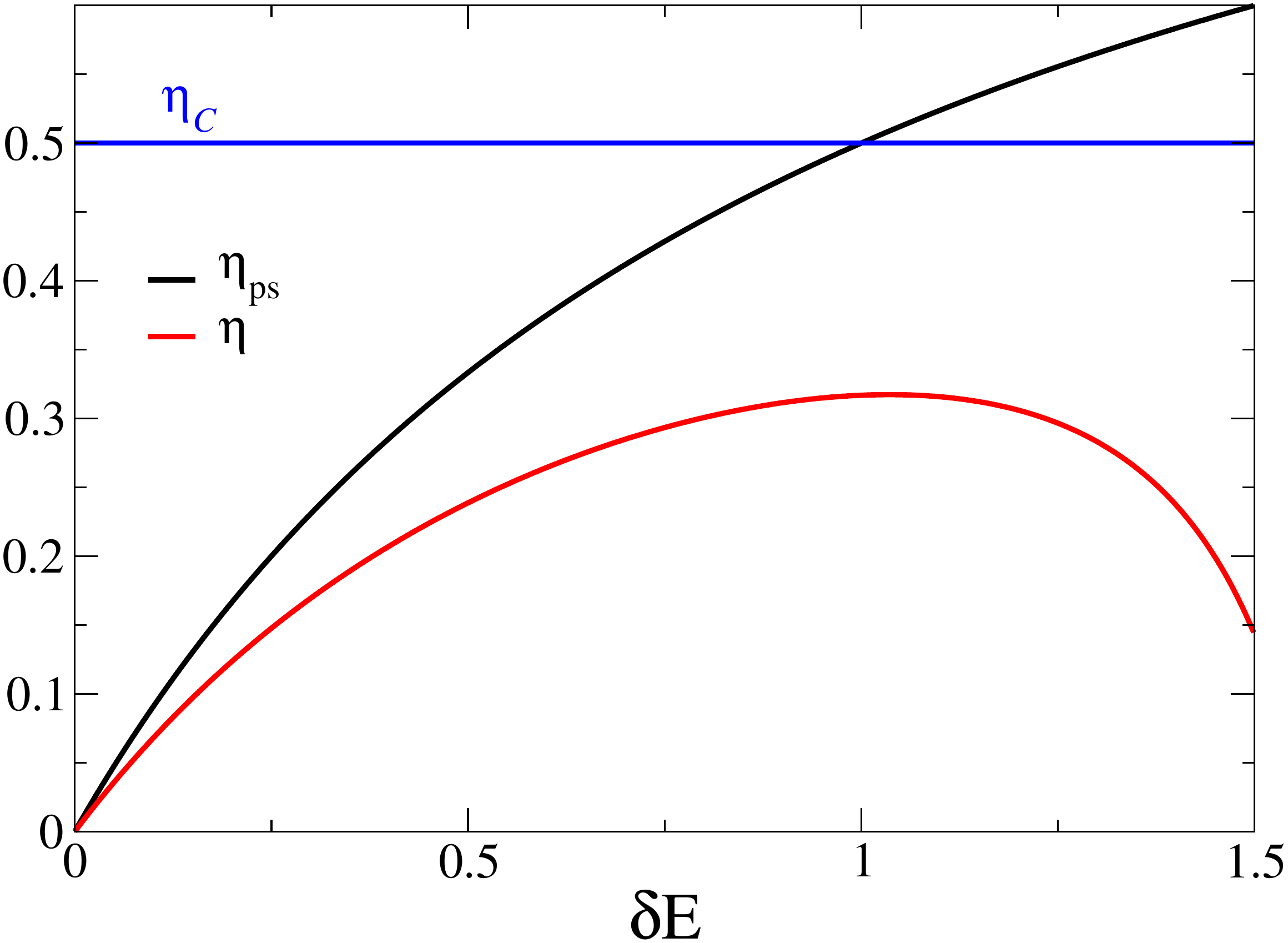}\label{fig6c}}\hspace{3mm}
\subfigure[]{\includegraphics[width=72mm]{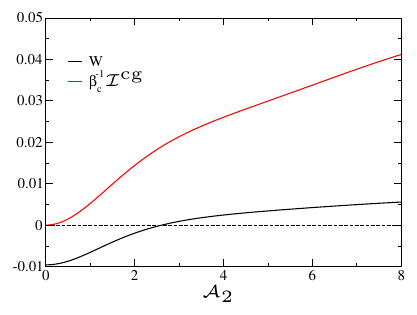}\label{fig6d}}
\vspace{-2mm}
\caption{(a) Depiction of the model for two interacting particles for the case the observable particle (in red) is at position $x=1$
and the hidden particle (in gray) is at position $a=-1$. The particles can jump to position $0$ with the rates given in the figure. 
(b) The quantities $\beta_c\eta_CQ_h-\beta_cW$, which can be negative for an active heat engine, and $\Delta S_{\textrm{ex}}^{\textrm{cg}}$ 
as a function of the inverse hot temperature $\beta_h$ for $\delta E= 0.2$ and $\A_2=10/3$. (c) Efficiency $\eta$ and pseudo-efficiency $\eta^\textrm{ps}$ as functions of of $\delta E$ and compared to the Carnot 
bound for $\beta_h=0.5$, $\A_2=10/3$. (d) Work $W$ and its upper bound $(\beta_c)^{-1}\hh^{\textrm{cg}}$ as functions of $\A_2$ for constant temperature $\beta_h=\beta_c$
and $\delta E=1.5$. For the three plots $\beta_c=1$, $\A_1=0$, $E=1$, $k=10$, and $\gamma=1$.}   
\label{fig6} 
\end{figure*}

To illustrate the applicability of our second law to the case with interactions between $x$ and $a$, we now consider a simple model for 
two interacting particles illustrated in Fig. \ref{fig6a}. The particles can be in three different positions. The observable particle has its position labeled by $x=-1,0,1$. 
The energy for this particle is $E(t)$ for $x=\pm 1$ and  zero for $x=0$.
The time-dependent energy is $E$ for the first part of the period and $E+\delta E$ for the third part of the period.
The position of the hidden particle is labeled by $a=-1,0,1$. The energy $E_a$ is zero for all $a$. This particle is driven in the clockwise direction 
by an affinity $\A$. Finally, the particles interact with each other through a hardcore potential. 

There are six possible states $i=a,x$. We label these states in the following way. State $i=1$ corresponds to  $x=-1$ and $a=0$.
State $i=2$ corresponds to  $x=-1$ and $a=1$. State $i=3$ corresponds to  $x=0$ and $a=-1$. State $i=4$ corresponds to  $x=0$ and $a=1$.  
State $i=5$ corresponds to  $x=1$ and $a=-1$. State $i=6$ corresponds to  $x=1$ and $a=0$. Using this labeling of the states, the transpose of the 
stochastic matrix for this model reads
\begin{equation}
\left(
\begin{array}{cccccc}
 -r_1  & \gamma e^{\frac{\beta  \A}{6}} & 0 & 0 & 0 & k  \\
 \gamma e^{-\frac{\beta  \A}{6}} & -r_2 & 0 & k  e^{\frac{\beta  E}{2}} & 0 & 0 \\
 0 & 0 & -r_3 & \gamma e^{-\frac{\beta  \A}{6}} & k  e^{-\frac{\beta  E}{2}} & 0 \\
 0 & k  e^{-\frac{\beta  E}{2}} & \gamma e^{\frac{\beta  \A}{6}} & -r_4 & 0 & 0 \\
 0 & 0 & k  e^{\frac{\beta  E}{2}} & 0 & -r_5 & \gamma e^{\frac{\beta  \A}{6}} \\
 k  & 0 & 0 & 0 & \gamma e^{-\frac{\beta  \A}{6}} & -r_6 \\
\end{array}
\right),
\end{equation}
where $k$ ($\gamma$) sets the time-scale for transitions of the observable (hidden) particle. The escape rates $r_i$ are the sum of non-diagonal elements of a row. 
To shorten the matrix we have omitted the time-dependence of $\A$, $E$, and $\beta$. For the first part of the period they are $\A_1$, $E$, and $\beta_c$ and 
for the third part of the period  they are $\A_2$, $E+\delta_E$, and $\beta_h$. The stationary distribution $P^S$ is the eigenvactor associated with the eigenvalue 
0 of this matrix. With this stationary distribution we obtain analytical expressions for all desired quantities defined in Sec. \ref{sec51}. 
The analytical expressions are too long to be displayed here.   

This model displays all the same features from the previous model, as illustrated in Fig. \ref{fig6}, which reiterates the fact that the second law 
for active heat engines is independent of the particular model. The efficiency in Fig. \ref{fig6c} is  $\eta\equiv  \frac{W}{Q_h+\hh^{\textrm{cg}}/(\beta_c-\beta_h)}$. 

\section{Numerical case study }
\label{secnew} 
\subsection{Relation to experiment }

The models of the active particle in a harmonic potential and of the two interacting particles have some similarities 
with the experiment in \cite{kris16}. There is an observable particle labeled by $x$. Furthermore,  heat and work can be evaluated by an expression that only depends 
on the observation of the variable $x$. The difference is in the hidden degrees of freedom. For the first model, this hidden degree of freedom 
is the sign of the force $F$. For the second model, the hidden degree of freedom is the position of the driven particle. For the experiment involving a colloidal
particle in a bacterial bath, the hidden degrees of freedom are more complex. They correspond to the positions and other variables that determine the state of the bacteria. 

However, the fact that the hidden degrees of freedom in the experiment are more complex does not present a challenge for our second law. The 
new term $\hh^{\textrm{cg}}$ can also be calculated by the sole observation of the variable $x$, no matter how complex the hidden degrees of freedom 
are. Our second law should be directly applicable to the experiment in \cite{kris16} and to future experiments with cyclic active heat engines.
To illustrate the generality of our second law we consider a more complex model than our analytical case studies. This model has several hidden 
degrees of freedom and is inspired by the experiment in \cite{kris16}.

\subsection{Model definition}

\begin{figure}
\includegraphics[width=8cm]{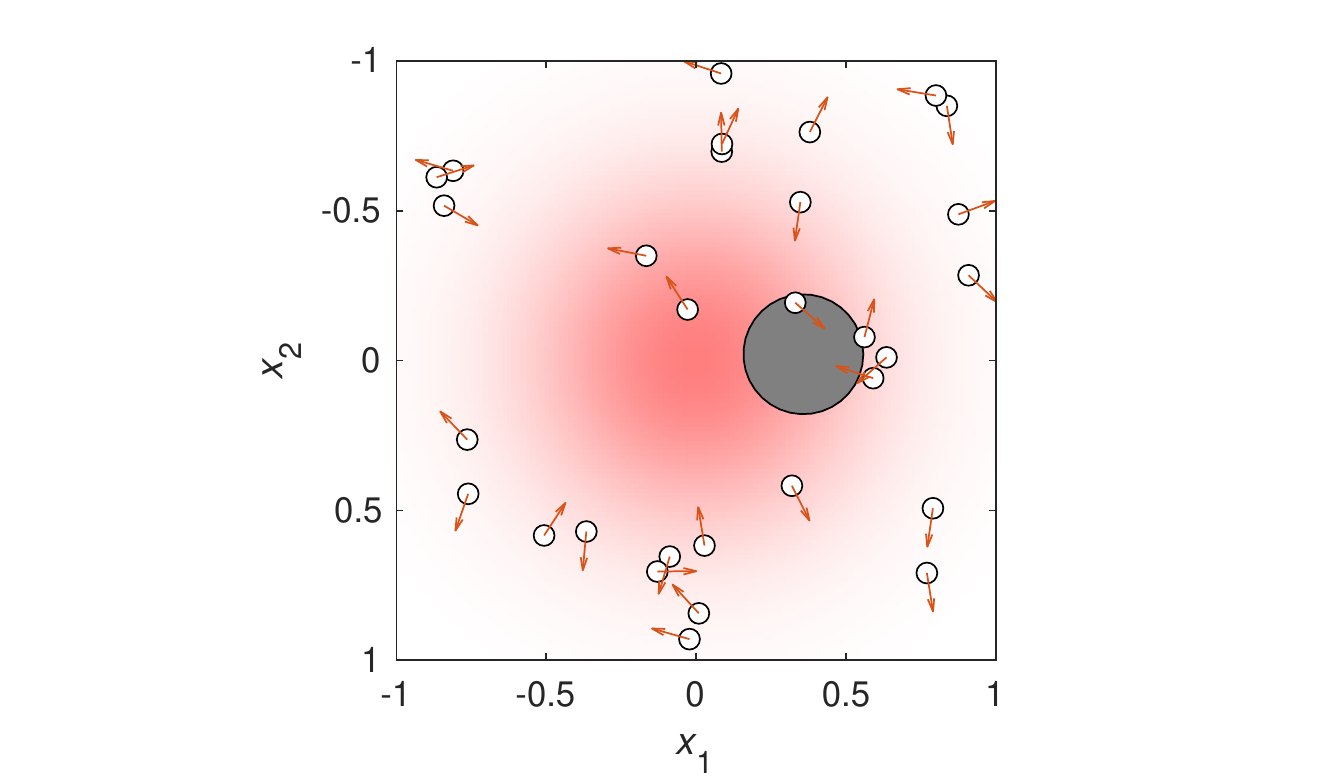}
\vspace{-2mm}
\caption{Setup for our numerical case study. A passive
    particle (grey) trapped in a harmonic potential (indicated in red)
    interacts with many active Brownian particles (white) that are self-propelled in
    randomly varying directions (orange arrows). 
}   
\label{fig7} 
\end{figure}

The two-dimensional model depicted in Fig. \ref{fig7} has a passive particle and many active particles that make the reservoir active.
These active particles correspond to the bacteria in the experiment. The passive particle has its position labeled by $\vec{x}=(x_1,x_2)$ and is subjected to a harmonic potential 
\begin{equation}
E(x)=\kappa x^2/2,
\end{equation}
where $\kappa$ is the stiffness of the potential and $x\equiv |\vec{x}|$. The active particles are not influenced by this harmonic potential and they do not interact with
each other. They all interact with the passive particle via the repulsive potential $V(|\vec{r}|)$, where  $\vec{r}$ is the distance to the 
passive particle. This potential is given by
\begin{equation}
V(|\vec{r}|)=
\left\{
	\begin{array}{ll}
		0  & \textrm{if }  |\vec{r}|>R \\
		\kappa_\mathrm{rep}(R-|\vec{r}|)^2/2 & \textrm{if } |\vec{r}|<R
	\end{array}
\right.,
\end{equation}
where $R$ is the radius of the passive particle, and $\kappa_\mathrm{rep}>0$ such that the interaction is repulsive. 

The passive particle obeys the Langevin equation 
\begin{equation}
  \dot{\vec{x}}=-\mu_p \nabla_{\vec{x}}E(x)-\mu_p\sum_m\nabla_{\vec{x}}V(|\vec{a}_m-\vec{x}|)+\vec\xi(t)
  \label{eq:langevinp}
\end{equation}
 where $\mu_p$ is the mobility of the passive particle. The term $\vec\xi(t)$ corresponds to Gaussian white noise with 
 intensity $\mu_p/\beta$. The sum over $m$ is over all active particles and $\vec{a}_m$ is the position of the active 
 paticle indexed by $m$.

The active particle labelled by $m$ obeys the Langevin equation
\begin{equation}
\dot{\vec{a}}_m=u\vec{n}(\phi_m)-\mu_a\nabla_{\vec{a}_m}V(|\vec{a}_m-\vec{x}|)+\vec\zeta_m(t),
\label{eq:langevina}
\end{equation}
where $\mu_a$ is the mobility of the active particles and $\vec{\zeta}_m(t)$ is a Gaussian white noise with intensity
$\mu_a/\beta$. For the first term on the left hand side, $u$ is the active speed, $\vec{n}(\phi_m)$ is a unit vector 
pointing in the direction of angle $\phi_m$. The angle $\phi_m$ undergoes Brownian diffusion with diffusion coefficient
$D_\mathrm{r}$ \cite{fily12,roma12}.   

For the active particles the vector $\vec{a}_m$ is constrained to the box shown in Fig. \ref{fig7} with periodic boundary conditions.
The passive particle is unconstrained. However, due to the harmonic potential $E(x)$ the passive particle rarely leaves this box. 

The protocol has the same four parts as the generic one shown in Fig. \ref{fig4}. 
Each part of the protocol takes the time $\tau/4$.
The temperature changes are the same as in all models analyzed here. 
The stiffness of the potential $E(x)$ switches from $\kappa_1$ 
to $\kappa_2$ from first to the second part and changes back from $\kappa_2$ to
$\kappa_1$ from the third to the fourth part. The velocity $u$ of the active 
particles also changes with the protocol. During the parts of the protocol where the 
inverse temperature is $\beta_c$ ($\beta_h$) this velocity is $u_c$ ($u_h$).
  
We here have set the parameters to the following values: the period is $\tau=20$; there are 30 active particles; the stiffness takes the values  
$\kappa_1=10$, $\kappa_2=15$; the inverse temperatures are $\beta_c^{-1}=0.1$, $\beta_h^{-1}=0.12$; the radius of the 
passive particle is $R=0.2$; the mobilities are  $\mu_p=0.05$ and $\mu_a=0.1$; the parameter of the repulsive potential is $\kappa_\mathrm{rep}=1000$;
the diffusion coefficient associated with the direction of the active speed is $D_\mathrm{r}=1$. The active speed is set as $u_h=u_0$ and 
$u_c=u_0/100$, where $u_0$ is parameter that we vary. For $u_0=0$ the heat engine is passive.

\subsection{Simulations and observables}

Simulations are performed by integrating the Langevin equations
\eqref{eq:langevinp} and \eqref{eq:langevina} using a simple stochastic Euler
scheme with a sufficiently small time step. We have used a time-step of 
$10^{-4}$. For each trajectory, we write stochastic 
work and heat as $w$ and $q_h$, respectively. The heat $Q_h$ and work $W$ are related to
the average of these quantities over stochastic trajectories.

From Eq. \eqref{eqW2}, the increments to the stochastic work $w$ happen at the two times of 
the period that the stiffness changes. For the change in the stiffness from first to the second part the increment 
in the work is $(\kappa_1-\kappa_2)\vec{x}^2/2$, where $\vec{x}$ is the position of the passive 
particle at the moment of the change.  For the second change in the stiffness from the third to the fourth part, the increment 
in the work is $(\kappa_2-\kappa_1)\vec{x}^2/2$.

From Eq. \eqref{eqjq2}, the increments to the stochastic heat $q_h$  happen when the passive particle changes position
during the third and fourth parts of the period, for which the inverse temperature is $\beta_h$. This increment can be formally
written as $dq_h=h(t)k(t)d(\vec{x}^2)/2$, where the function $h(t)$ in Eq. \eqref{eqbeta} is $1$ if $\beta(t)=\beta_h$ and 
$0$ if $\beta(t)=\beta_c$.

Estimates for the average heat $Q_h$ and work $W$ per cycle then follow from the final value of $q_h$ and $w$ divided by the number of cycles
performed in a long trajectory (we average over 5000 cycles).

\begin{figure}
\includegraphics[width=9cm]{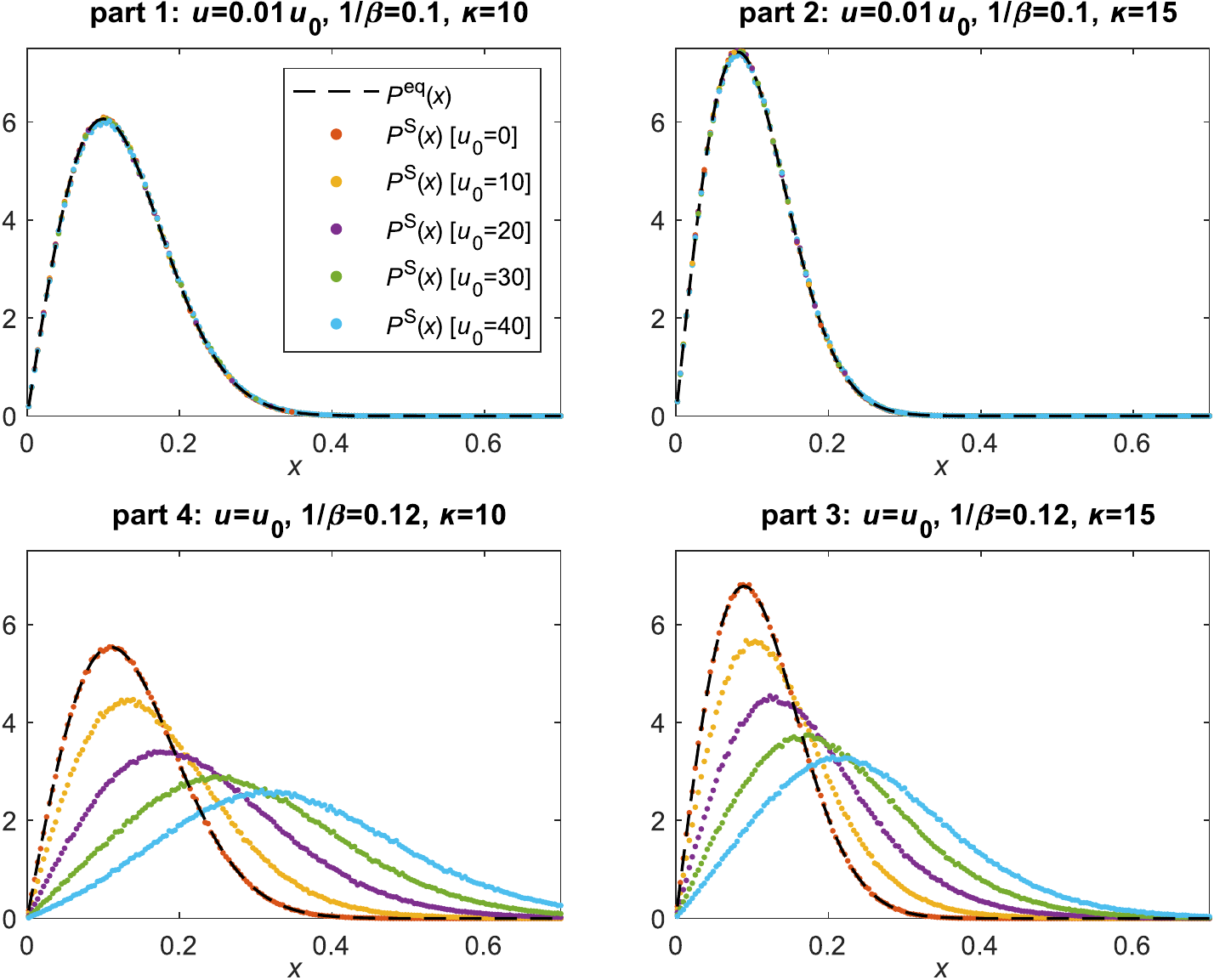}
\vspace{-2mm}
\caption{Stationary distributions  sampled with fixed parameters according to each of the four steps of the
    protocol. The reference active speed $u_0$ corresponds to different curves indicated by color. The
    equilibrium distribution is shown as a dashed black curve. For parts 1 and 2, the stationary distribution are 
    closer to the equilibrium distribution due to the small driving speed $u$.
}   
\label{fig8} 
\end{figure}

The calculation of $\mathcal{I}^\mathrm{cg}$ is more involved. From Eq. \eqref{eqhhcg2}, to calculate the increment associated
with this quantity, we need the equilibrium distribution 
\begin{equation}
P^{\mathrm{eq}}(x)=\beta kx\exp(-\beta kx^2/2)
\end{equation}
and the accompanying distribution $P^S$. 
Since the protocol has four parts, there are four stationary distributions $P^{S,l}(x)$. To sample these distributions we  run four separate simulations where $\kappa$, $\beta$ and $u$ remain constant.
The results for these distributions are shown in Fig. \ref{fig8}. This method of sampling $P^{S,l}(x)$ from realizations with fixed parameters that reach a steady state
can also be used in experiments  for which  this probability is not known. 

The stochastic quantity corresponding to $\mathcal{I}^\mathrm{cg}$ can now be calculated in similar way to the stochastic work $w$. Whenever there is a change 
from part $l$ to part $l+1$ of a cycle and the passive particle is in position $\vec{x}$,  we add up the increment  
$\ln\frac{P^{S,l}(x)P^{\mathrm{eq},l+1}(x)}{P^{\mathrm{eq},l}(x)P^{S,l+1}(x)}$. 
To get an estimate of the average $\mathcal{I}^\mathrm{cg}$ we divide the total sum of all increments during the trajectory by the number of cycles.
 
We reiterate that our estimates of $W$, $Q_h$ and $\mathcal{I}^\mathrm{cg}$ only depend on the monitoring of the position of the passive 
particle $\vec{x}$. This scheme can also be applied to an experiment where one does not have access to the hidden active degrees 
of freedom, such as the experiment in \cite{kris16}.

\subsection{Results}

\begin{figure}
\subfigure[]{\includegraphics[width=8cm]{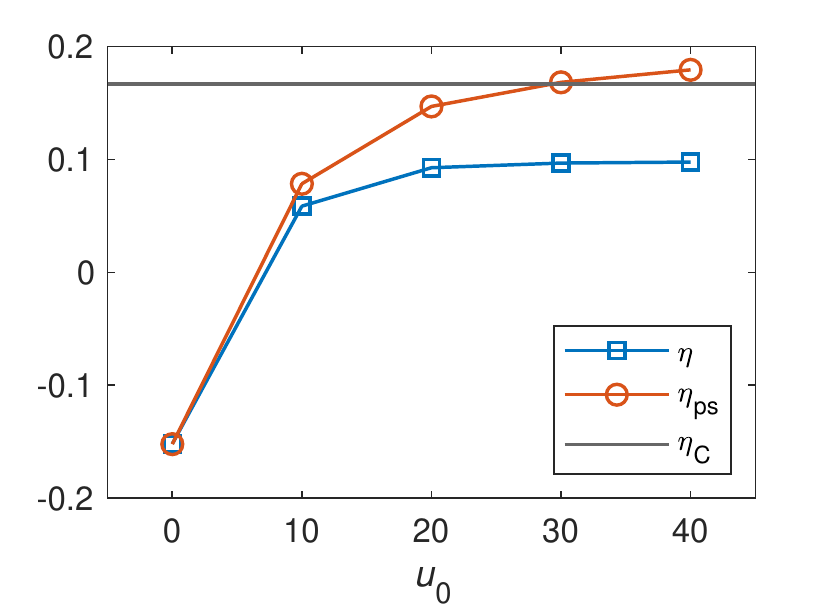}\label{fig9a}}
\subfigure[]{\includegraphics[width=8cm]{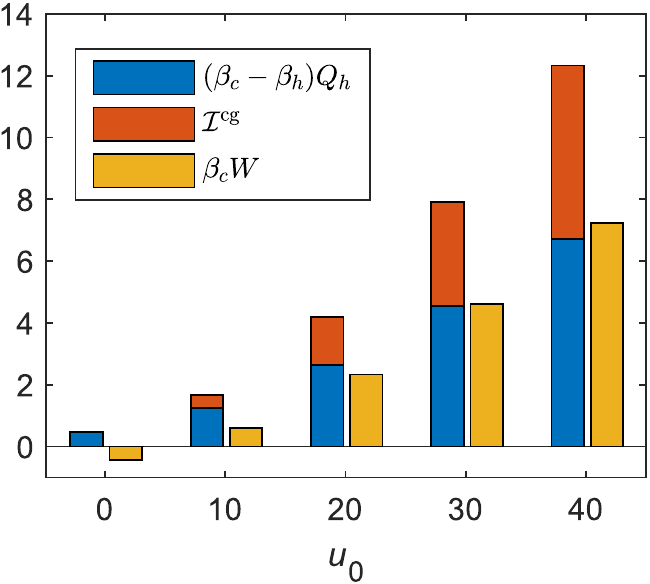}\label{fig9b}}
\vspace{-2mm}
\caption{Results for numerical model. (a) Efficiency (blue squares)
    and pseudo efficiency (red circles), as a function of the reference active
    speed $u_0=u_h$. Error bars calculated from the standard deviation with our data are not larger than the size of the points in the figure.
     (b) The quantities $\beta_cW$, $\mathcal{I}^\mathrm{cg}$, and $(\beta_c-\beta_h)Q_c$.
}   
\label{fig9} 
\end{figure}

The results for for the numerical model are shown in Fig. \ref{fig9}. They are plotted as a function of $u_0$,
which parameterize the active speed. For $u_0= 0$ the heat engine is passive. We observe that 
for increasing $u_0$ the active heat engine becomes more efficient. In fact, for our chosen parameters, 
activity allows the system to operate as a heat engine since for $u_0=0$ the extracted work $W$ is negative. The 
pseudo efficiency $\eta_{\textrm{ps}}\equiv W/Q_h$ does go beyond the Carnot bound while the efficiency $\eta\equiv  \frac{W}{Q_h+\hh^{\textrm{cg}}/(\beta_c-\beta_h)}$
stays below the bound, as predicted by our second law. 

As shown in Fig. \ref{fig8}, with increasing active speed $u_0$ the stationary distribution
departs further from the equilibrium distribution. In Fig. \ref{fig9} we demonstrate that the 
term $\mathcal{I}^\mathrm{cg}$, which quantifies the difference between these distributions, increases, from about $30\,\%$ of the value of
$(\beta_c-\beta_h)Q_h$ for $u_0=10$ to $80\,\%$ for $u_0=40$. When the engine
is highly active, $\mathcal{I}^\mathrm{cg}$ acts as an important resource for the extraction of work.

Related models have been analyzed in Refs.~\cite{zaki17,holu20}, using Langevin equations for the passive particle  with a noise that 
is not the standard Gaussian white noise. This noise effectively captures the influence of the active particles that interact with the 
passive particle. Both these studies use the concept of an effective temperature. An advantage of our approach is that we have no need 
to invoke an effective temperature, our second law inequality contains {\sl bona fide} temperatures.

\section{Conclusion}
\label{sec6}

The second law of thermodynamics determines which processes are possible. For well known passive heat engines, 
this statement is expressed as the Carnot bound on the efficiency. We here derived the appropriate statement of the 
second law for cyclic active heat engines. Unlike a passive heat engine, the second law for an active heat engine includes an information-theoretic term, 
which allows active heat engines to perform tasks beyond the ones that can be performed by passive heat engines, such as work 
extraction at constant temperature.

Our second law for cyclic active heat engines has three main merits. First, it does not include the dissipation due to the 
hidden degrees of freedom. Second, it is expressed in terms of quantities that can be evaluated by the sole
monitoring of the observable degrees of freedom. Third, it is general, as it only requires the mild assumption that 
at some level of description the observable degrees of freedom together with the hidden degrees of freedom follow 
Markovian dynamics. Furthermore, the system corresponding to the observable degrees of freedom does not have to be 
a single colloidal particle, which is the case for most models analyzed in the literature, but it can be any 
arbitrary collection of possibly interacting degrees of freedom.

From the perspective of stochastic thermodynamics, we have unveiled a central application of excess entropy in 
periodically driven systems. While this quantity has been known to fulfill a fluctuation theorem we here 
show that it provides the correct statement of the second law for cyclic active heat engines. Beyond the known 
excess entropy we introduced a coarse-grained excess entropy that was key for the derivation of the second law 
in terms of quantities that only depend on the observable degrees of freedom. Interestingly, the second law for active heat engines retains its structure under coarse-graining.
This feature highlights the power of stochastic thermodynamics, which also applies to an active reservoir with hidden dissipative degrees of freedom that leads to 
non-Markovian dynamics.

Active matter is an important modern concept in physics. Quite generally, if we drive a system immersed in such active matter 
with an external periodic protocol with the objective of completing some task such as work extraction, the second law derived here 
constitutes a general bound on whether the completion of this task is possible. Therefore, our results are relevant for 
future experiments with active heat engines. From the theoretical side, the second law derived here opens up the possibility 
to develop a linear response theory for active heat engines, it is central for problems such as the optimization of work and efficiency, 
and it should be useful for future models of, possibly interacting, active heat engines.

\appendix

\section{Fluctuation theorem for coarse-grained excess entropy}

To prove the second law for coarse grained excess entropy we must define quantities as functionals 
of stochastic trajectories from time $t=0$ up to time $t=T$. The time $t$ is discretized with time-intervals $\delta t$ to simplify the expression 
for the probability of a trajectory, the continuous time limit is recovered for $\delta t\to 0$. For discrete time, the transition probability from 
$i$ to $j$ is $M_{ij}(t)=\delta tk_{ij}(t)$ for $i\neq j$ and $M_{ii}(t)=1-\sum_{j\neq i}\delta tk_{ij}(t)$.
For the proof in this appendix the time-dependence of the transition rates is arbitrary, there is no need to assume a periodic time-dependence. For the calculations below, a key quantity is the dual transition rate 
$k^{\dagger}_{ij}(t)= k_{ji}(t)P^S_j(t)/P^S_i(t)$ (the same equation applies for transition probabilities $M^{\dagger}_{ij}(t)$), where $P^S_i(t)$ is the stationary probability for fixed rates at time $t$, i.e., they are the solution of Eq. \eqref{eqProb}. 

The probability of a trajectory $\Gamma= i_0\to i_1\ldots i_{N-1}\to i_N$, where $N=T/\delta t$, is given by
 \begin{equation}
\pr(\Gamma)=P_{i_0}(0)\prod_{n=0}^{N-1}M_{i_ni_{n+1}}(n\delta t).
\label{eqweight}
\end{equation}
The probability of the reversed trajectory with dual dynamics (without reversing the time-dependence of the transition rates) is
 \begin{equation}
\pr^\dagger(\Gamma)=P_{i_N}(N\delta t)\prod_{n=0}^{N-1}M^{\dagger}_{i_{n+1}i_n}(n\delta t).
\end{equation}
The trajectory dependent excess entropy change is defined as 
\begin{equation}
\Delta S_{\textrm{ex}}(\Gamma)\equiv \sum_{n=0}^{N-1}\ln\frac{M_{i_ni_{n+1}}(n\delta t)}{M^{\dagger}_{i_{n+1}i_n}(n\delta t)}= \sum_{n=0}^{N-1}\ln\frac{P^S_{i_{n+1}}(n\delta t)}{P^S_{i_{n}}(n\delta t)}.
\end{equation}
Furthermore, the system entropy change is
\begin{equation}
\Delta S_{\textrm{sys}}(\Gamma)\equiv \ln\frac{P_{i_0}(0)}{P_{i_N}(N\delta t)}.
\end{equation}
Therefore, the sum of these two terms gives 
\begin{equation}
\Delta S_{\textrm{ex}}(\Gamma)+\Delta S_{\textrm{sys}}(\Gamma)= \ln\frac{\pr(\Gamma)}{\pr^*(\Gamma)}.
\end{equation}
From this formula it is easy to show that the excess entropy fulfills a fluctuation theorem \cite{seif12}. A useful equation is   
\begin{equation}
\langle \Delta S_{\textrm{ex}}+\Delta S_{\textrm{sys}}\rangle= D_{KL}(\pr||\pr^\dagger)\ge 0,
\label{eqdkl1}
\end{equation}
where the brackets indicate an average over stochastic trajectories and the Kullback-Leibler distance here is over the space of trajectories $\Gamma$. If the transition rates are time-periodic with period 
$\tau$, the final time is $T=\tau$, and at the starting time $t=0$ the system has already reached the asymptotic time-invariant probability,  this inequality becomes the second law used 
in Eq. \eqref{eqexc}. Note that the average change of system entropy is zero if the system entropy is time-periodic.

To define the coarse-grained excess entropy we write $i=(x,a)$, where $x$ is the observable variable and $a$ is hidden. The dynamics for the coarse-grained variable $x$ is not Markovian. The probability of 
a trajectory in the space of trajectories of $x$ is defined as 
\begin{equation}
\pr(\Gamma^{\textrm{cg}})=\sum_{a_0,a_1,\ldots,a_N}\pr(\Gamma),
\end{equation}    
where $\Gamma^{\textrm{cg}}= x_0\to x_1\ldots x_{N-1}\to x_N$.
In order to express this probability in a form similar to Eq. \eqref{eqweight} we use the quantity $\mathcal{M}_{xx'}(t)=\sum_{a,a'}P_{a|x}(t)M_{x,a;x',a'}(t)$, which leads to the expression
\begin{equation}
\pr(\Gamma^{\textrm{cg}})=P_{x_0}(0)\prod_{n=0}^{N-1}\mathcal{M}_{x_nx_{n+1}}(n\delta t).
\label{eqweight2}
\end{equation}
For this expression we used $P_{x_0,a_0}(0)=P_{x_0}(0)P_{a_0|x_0}(0)$ and 
\begin{align}
\sum_{a_n}P_{a_n|x_n}(n\delta t)M_{x_n,a_n;x_{n+1},a_{n+1}}(n\delta t)=\nonumber\\
 \mathcal{M}_{x_nx_{n+1}}(n\delta t)P_{a_{n+1}|x_{n+1}}((n+1)\delta t).
\end{align}
The same formula is valid for the reversed trajectory with dual rates, i.e.,
\begin{equation}
\pr^\dagger(\Gamma^{\textrm{cg}})=P_{x_N}(N\delta t)\prod_{n=0}^{N-1}\mathcal{M}^{\dagger}_{x_nx_{n+1}}(n\delta t),
\label{eqweight3}
\end{equation}
where $\mathcal{M}^{\dagger}_{xx'}(t)=\sum_{a,a'}P_{a|x}(t)M^{\dagger}_{x,a;x',a'}(t)$  .

The coarse-grained excess entropy is defined as 
\begin{equation}
\Delta S_{\textrm{ex}}^{\textrm{cg}}(\Gamma)\equiv \sum_{n=0}^{N-1}\ln\frac{\mathcal{M}_{x_nx_{n+1}}(n\delta t)}{\mathcal{M}^{\dagger}_{x_{n+1}x_n}(n\delta t)}= \sum_{n=0}^{N-1}\ln\frac{P^S_{x_{n+1}}(n\delta t)}{P^S_{x_{n}}(n\delta t)}.
\end{equation}
The coarse-grained system entropy change is
\begin{equation}
\Delta S_{\textrm{sys}}^{\textrm{cg}}(\Gamma)\equiv \ln\frac{P_{x_0}(0)}{P_{x_N}(N\delta t)}.
\end{equation}    
The sum of both terms gives
\begin{equation}
\Delta S_{\textrm{ex}}^{\textrm{cg}}(\Gamma)+\Delta S_{\textrm{sys}}^{\textrm{cg}}(\Gamma)= \ln\frac{\pr(\Gamma^{\textrm{cg}})}{\pr^\dagger(\Gamma^{\textrm{cg}})}.
\label{eqft1}
\end{equation}
Hence, 
\begin{equation}
\langle \Delta S_{\textrm{ex}}^{\textrm{cg}}+\Delta S_{\textrm{sys}}^{\textrm{cg}}\rangle= D_{KL}^{\textrm{cg}}(\pr||\pr^\dagger)\ge 0,
\label{eqdkl2}
\end{equation}
where we have used the superscript cg in the Kullback-Leibler distance to indicate that 
we are considering probabilities in the space of trajectories $\Gamma^{\textrm{cg}}$. 

For transition rates that are time-periodic with period $\tau$, if the time interval is $T=\tau$ and 
if at the starting time $t=0$ the system has already reached the asymptotic time-invariant probability, the inequality in Eq. \eqref{eqdkl2} becomes the coarse-grained second law used in  
Eq. \eqref{eqeqexcg3}. The average  system entropy is periodic, and, therefore, the average system entropy change in a period is zero. 

The average excess entropy is larger than the average coarse-grained excess entropy, since 
\begin{equation}
D_{KL}(\pr||\pr^\dagger)\ge D_{KL}^{\textrm{cg}}(\pr||\pr^\dagger),
\label{eqdkl3}
\end{equation}
which implies the inequality in Eq. \eqref{eqexcine}. A similar inequality for the standard entropy production has been used in \cite{rold12}.

Finally, we can prove a fluctuation theorem for the coarse-grained excess entropy using relation \eqref{eqft1}, 
\begin{equation}
\langle \exp\left(-\Delta S_{\textrm{ex}}^{\textrm{cg}}-\Delta S_{\textrm{sys}}^{\textrm{cg}}\right)\rangle^{\textrm{cg}}= \sum_{\Gamma^{\textrm{cg}}}\frac{\pr^\dagger(\Gamma^{\textrm{cg}})}{\pr(\Gamma^{\textrm{cg}})}\pr(\Gamma^{\textrm{cg}})=1.
\end{equation}


%

\end{document}